\makeatletter

\newcommand{\Rmnum}[1]{\expandafter\@slowromancap\romannumeral #1@}
\makeatother

\documentclass[10pt,conference,twocolumn]{IEEEtran}
\usepackage{array}
\usepackage{amsmath,graphicx,epsfig,float,subfigure,times,latexsym,verbatim,mathrsfs,amssymb,textcomp,txfonts,color,setspace}
\usepackage{CJK}
\usepackage{algorithm}
\usepackage{algorithmic}
\usepackage{url}

\ifCLASSINFOpdf
\else
\fi
\begin{document}

\title{Increasing Indoor Spectrum Sharing Capacity using Smart Reflect-Array}

\author{\IEEEauthorblockN{{Xin Tan}, {Zhi Sun}, {Josep M. Jornet}, and {Dimitris Pados}}
\IEEEauthorblockA{Department of Electrical Engineering\\State University of New York at Buffalo, Buffalo, NY 14260\\
E-mail:
\{xtan3, zhisun, jmjornet, pados\}@buffalo.edu.}
}

\maketitle

\begin{abstract}
The radio frequency (RF) spectrum becomes overly crowded in some indoor environments due to the high density of users and bandwidth demands. To accommodate the tremendous wireless data demands, efficient spectrum-sharing approaches are highly desired. To this end, this paper introduces a new spectrum sharing solution for indoor environments based on the usage of a reconfigurable reflect-array in the middle of the wireless channel. By optimally controlling the phase shift of each element on the reflect-array, the useful signals for each transmission pair can be enhanced while the interferences can be canceled. As a result, multiple wireless users in the same room can access the same spectrum band at the same time without interfering each other. Hence, the network capacity can be dramatically increased. To prove the feasibility of the proposed solution, an experimental testbed is first developed and evaluated. Then, the effects of the reflect-array on transport capacity of the indoor wireless networks are investigated. Through experiments, theoretical deduction, and simulations, this paper demonstrates that significantly higher spectrum-spatial efficiency can be achieved by using the smart reflect-array without any modification of the hardware and software in the users' devices.
\end{abstract}


%
\IEEEpeerreviewmaketitle

\linespread{1}

\section{Introduction}
The radio frequency (RF) spectrum is becoming overly crowded due to an exponential growth in the number of applications and users that require high-speed wireless communication, anywhere, anytime \cite{1, 2, 3, 4, 5, 6, 7}. The situation just gets worse in indoor environments, such as conference halls or shopping malls, where both the user density and the bandwidth demands are tremendous. Moreover, The indoor RF spectrum is crowdedly occupied by a plethora of coexisting wireless services, including cellular networks, WiFi networks, Bluetooth systems, Wireless Sensor Networks or the Internet of Things, to name a few. To accommodate the tremendous wireless data demands from high density of users of different services, high-efficiency spectrum-sharing approaches are highly desired.

Currently, the cognitive radio \cite{6, 7, 8, congnitive1} and adaptive beamforming \cite{beamforming1, beamforming2, beamforming3} are two main dynamic spectrum access solutions that have drawn most attentions. On the one hand, cognitive radio techniques enable unlicensed wireless users to share channels with licensed users that are already using an assigned spectrum \cite{6}. Although cognitive radio achieves better spectrum efficiency when the licensed users do not access the band frequently (such as TV white space), it does not help when all users are very active with high density, especially in the aforementioned indoor environments. More importantly, cognitive radio techniques require each wireless device to be able to scan a wide range of frequencies to identify spectrum holes and then lock to that frequency, which necessitates expensive transceivers, antennas, and processors that are not available in most existing devices, if not all. On the other hand, usually beamforming techniques require smart antenna systems to adaptively focus the transmission and reception of wireless signals \cite{6, beamforming3, beamforming4}. To achieve high spatial resolution to differentiate multiple simultaneous transmissions in the crowded indoor environments, each wireless device needs to be equipped with very large array of antenna elements, which is impossible for the current and future portable, wearable, or even smaller devices.



\begin{figure}
  \centering
  \includegraphics[width=3.5in]{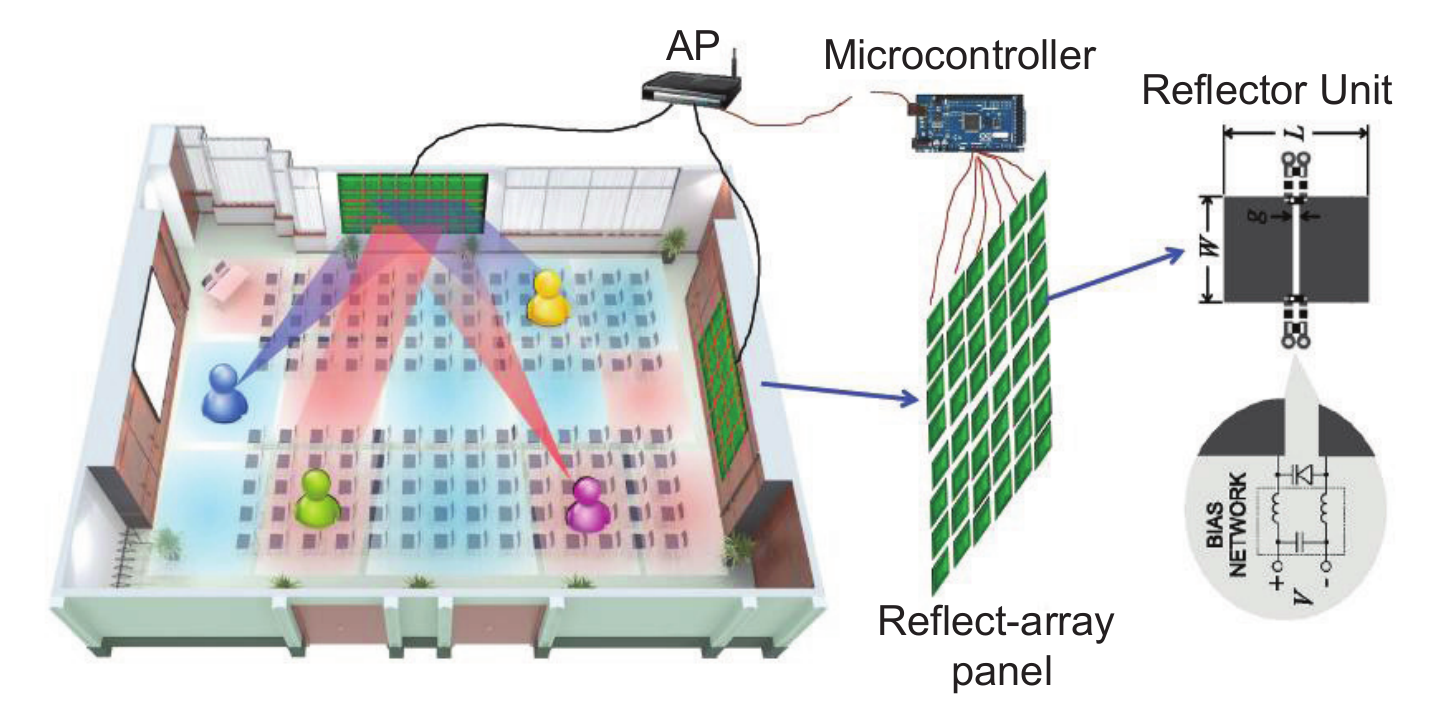}
    \vspace{-10pt}
  \caption{System architecture of spectrum sharing using smart reflect-array.}
  \vspace{-10pt}
  \label{fig.0}
\end{figure}

To address the aforementioned problems, in this paper, we propose a new spectrum sharing technique based on smart reflect-arrays to improve indoor network capacity for literally any existing wireless services without any modification of the hardware and software in the devices.
As shown in Fig. \ref{fig.0}, the smart reflect-array panels are hung on the walls in the indoor environment. Although the reflect-array does not buffer or process any incoming signals, it can change the phase of the reflected wireless signal. By optimally controlling the phase shift of each element on the reflect-array, the useful signals for each transmission pair can be enhanced while the interferences can be canceled. As a result, multiple wireless users in the same room can access the same spectrum band at the same time without interfering each other. To prove the feasibility of the proposed solution, an experimental testbed is first developed and evaluated. Then, the effects of the reflect-array on transport capacity of the indoor wireless networks is investigated. Through experiments, theoretical deduction, and simulations, this paper demonstrates that significantly higher spectrum-spatial efficiency can be achieved by using the smart reflect-array without any modification of the hardware and software in the users' devices.

The remainder of this paper is organized as follows. The system design and proof experiment are introduced in Section II. Then, the transport capacity, including theoretical upper bounds and achievable bounds for arbitrary networks, is derived and analyzed in Section III. Numerical analysis is presented in Section IV. Finally, the paper is concluded in Section V.

\section{System Architecture and Proof-of-Concept Experiment}
In this section, we first present the system architecture of the new reflect-array-based spectrum sharing solution. Then, an experimental testbed is designed and implemented to prove the feasibility of the proposed solution.

\subsection{System Architecture}

The architecture of the proposed system is illustrated in Fig. \ref{fig.0}. There are two pairs of wireless users in a conference room, whose devices can adopt any existing or future wireless services. The smart reflect-arrays hung on the walls can effectively change the signal propagations of any wireless transmission by tuning the electromagnetic response (phase shift) of each reflector unit on the panel. Hence, the wireless signal from either transmitter can be spatially modulated and projected to arbitrary regions while not interfering other regions. As a result, each receiver clearly hears from the expected transmitter as if only such transmitter accesses the spectrum, while actually there can be many other wireless users and services simultaneously use the same frequency band. The reflect-array panel actually reconfigure the signal propagation. The spatial distribution of the signal strength from different transmitters forms a chessboard of high resolution regions, each of which is private for only one wireless transmission. Since it is the reflect-array that manipulates the spatial modulation, the users can use any type of wireless devices and wireless services without any change in hardware or software.

Different from existing MIMO, beamforming, or active relay techniques, the proposed smart reflect-array moderates the spatial distribution of multiple wireless transmissions in a passive way. As long an EM wave-carried wireless transmission exists in the indoor environment, no matter where it comes from (e.g., cell phone, laptop, bluetooth speaker, smart home sensors, or cleaning robots), the smart reflect-array(s) can reconfigure the spatial distribution of the wireless energy due to such transmission. Moreover, different from existing beamforming mechanisms, the proposed system achieves the spatial diversity in the middle of wireless transmissions, neither in the transmitter nor in the receiver. This property further guarantees the compatibility to all possible wireless systems.

\subsection{Experimental Testbed Designed and Implemented}

To validate the feasibility of the proposed system, we develop a experiment testbed. Since we expect to have a flexible control of the electromagnetic response on reflect-array, it is necessary to optimally design the reflect-array panel and its peripheral circuits. For the reflect-array design, the basic idea is that by loading the microstrip patches with electronically-controlled capacitors, the resonant frequency of each reflector unit can be changed to increase the usable frequency range \cite{reflectarray}. More importantly, since the signals are required to be efficiently reflected on the reflect-array, the patches should be designed to have a satisfying reflection coefficient.

\begin{figure}
\centering
\subfigure[The smart reflect-array design.]{
\includegraphics[width=1.7in]{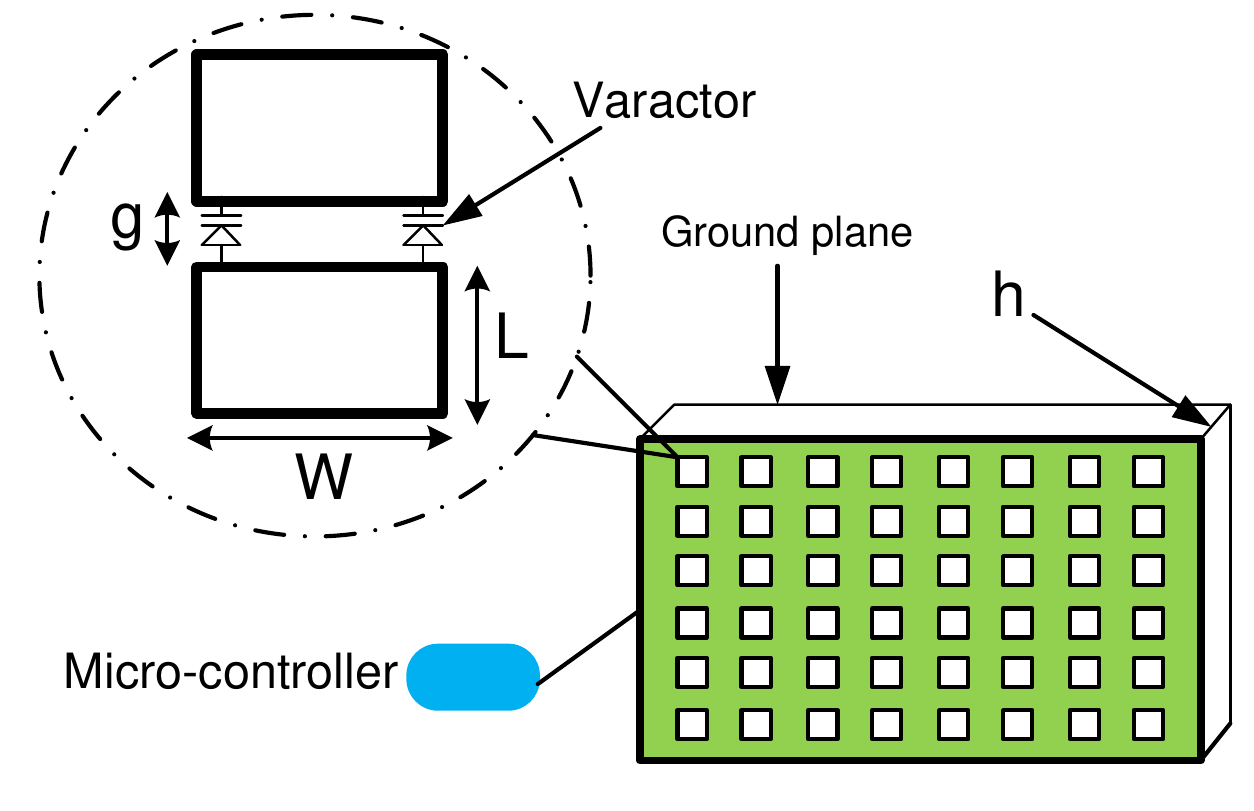}
\label{fig.design}} \quad \subfigure[The fabricated reflect-array.]{
\includegraphics[width=1.4in]{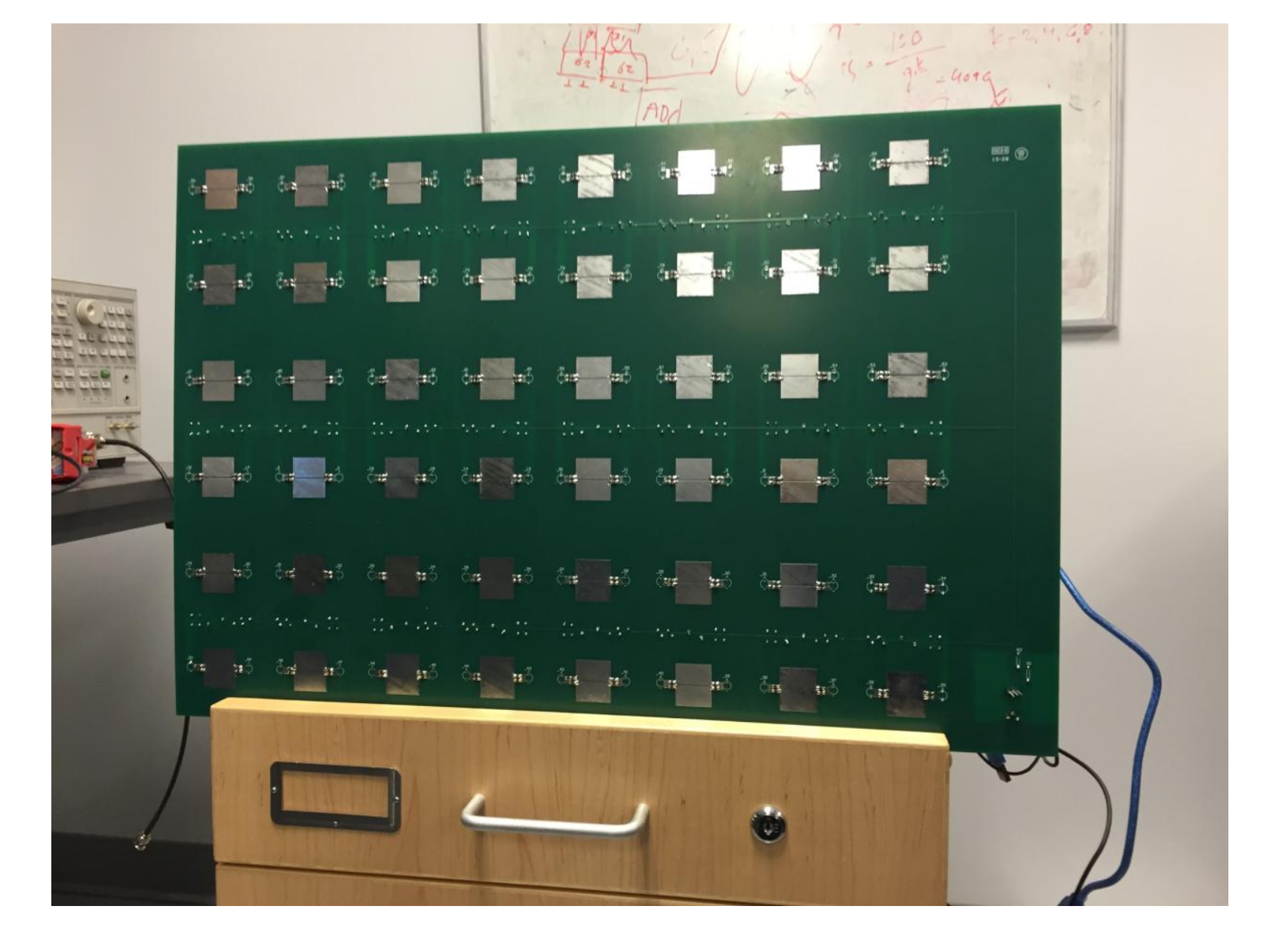}
\label{fig.view}}
\caption{Testbed design of the smart reflect-array.}\label{fig.designandview}
\end{figure}

\begin{figure}
\centering
\subfigure[RCS of the reflector unit]{
\includegraphics[width=1.6in]{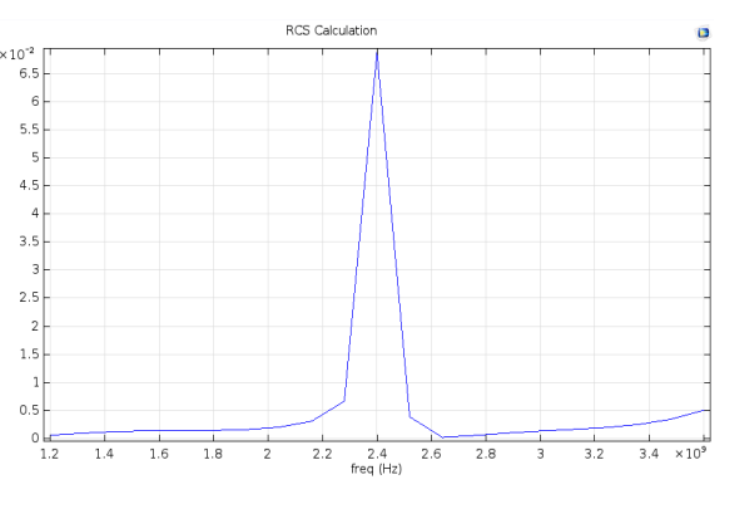}
\label{fig.RCS}}
\subfigure[E-field on the reflector unit]{
\includegraphics[width=1.6in]{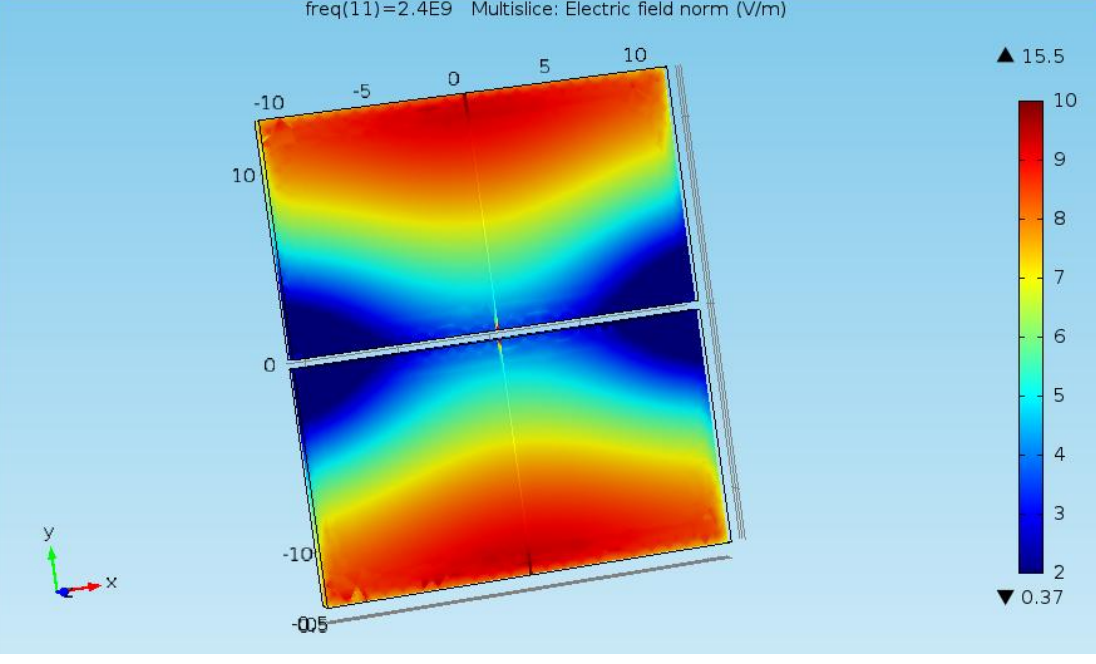}
\label{fig.energy}}
\caption{COMSOL simulation results.}\label{fig.simulation}
\end{figure}

In this design, the reflect-array is used to work at an operating frequency of 2.4 GHz, which is suitable for Wifi service in indoor environments. We design rectangle-structure patches as introduced in \cite{reflectarray} for the reflector units as shown in Fig. \ref{fig.design}, where the dimensions of the patch are developed as $W=25$ $mm$, $L=12.5$ $mm$, $g=0.5$ $mm$. The distance between the patch and ground plane is $h=1.5$$mm$. The relative dielectric constant $\varepsilon_r=4.5$, which can be realized by most of the PCB fabrications. We design totally $6 \times 8 = 48$ reflector units and each reflector is controlled by a bias voltage to tune the varactors ($0.6 - 8$ $pF$) for changing the capacitance. A view of developed smart reflect-array is shown in Fig. \ref{fig.view}. Simulations in Fig. \ref{fig.simulation} show the electromagnetic response the reflector unit by COMSOL. As shown in Fig. \ref{fig.RCS} , by using an operating frequency of 2.4 GHz, the designed patch get the maximum radar cross section (RCS), which means the reflection is optimized at such operating frequency. The energy distribution on the patch shown in Fig. \ref{fig.energy} further demonstrates that the resonance can be obtained at 2.4 GHz.

\begin{figure}
\centering
\subfigure[MEGA2560 micro-controller]{
\includegraphics[width=1.45in]{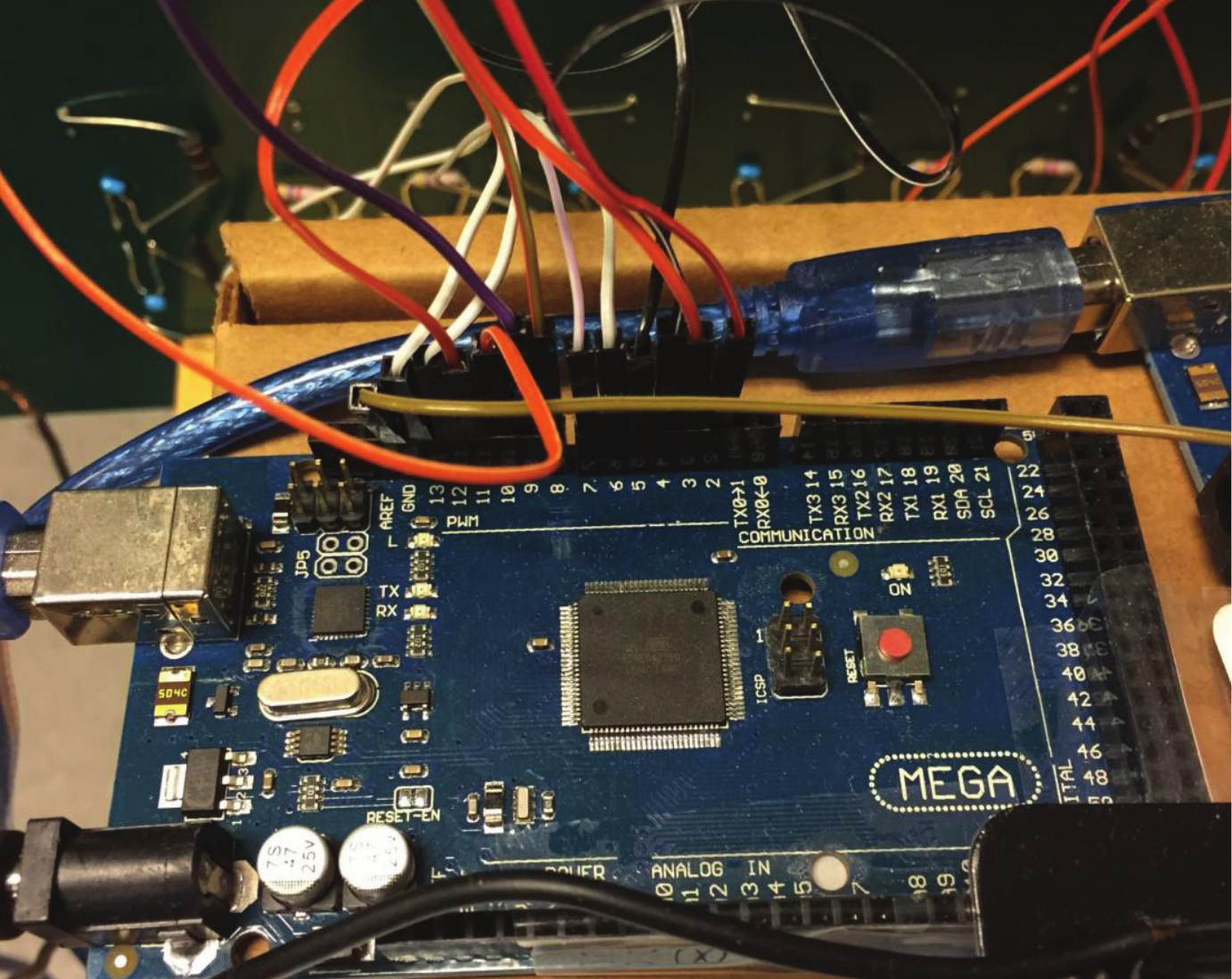}
\label{fig.02a}} \subfigure[RC low-pass filter]{
\includegraphics[width=1.5in]{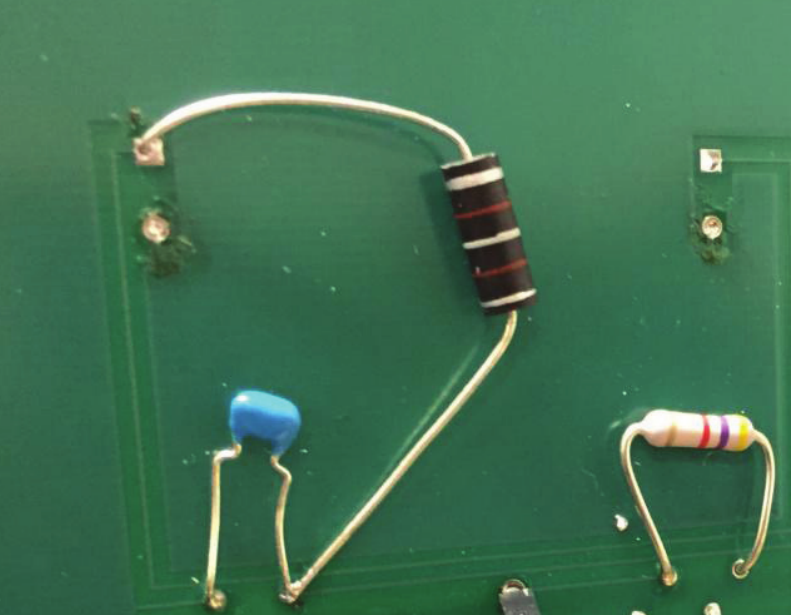}
\label{fig.02b}}
\caption{The micro-controller and RC low-pass filter used to control the reflect-array.}\label{fig.02}
\end{figure}

In this reflect-array, we use MEGA2560 micro-controllers as Fig. \ref{fig.02a} to generate PWM signals to control the reflectors. Since totally $48$ reflectors need to be independently controlled, $4$ micro-controllers are used that each one outputs 12 independent PWM signals. RC low-pass filter is designed as Fig. \ref{fig.02b} to convert the PWM signal to a certain bias voltage from 0 to 5 volts. Therefore, the reflectors become flexible to change the electromagnetic response to control the signals reflected on them.

\begin{figure}
\centering
\subfigure[Experiment illustration.]{
\includegraphics[width=2in]{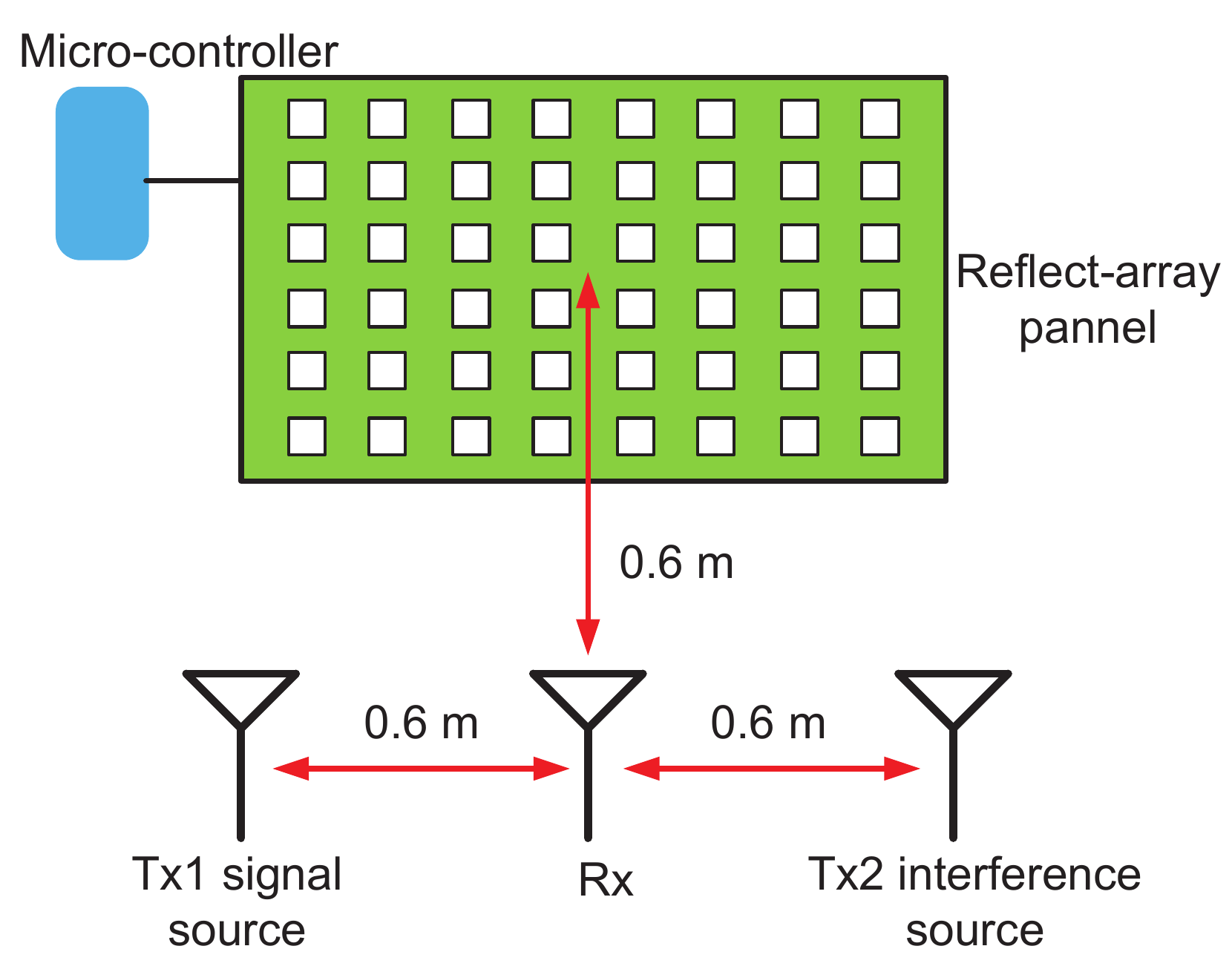}
\label{fig.01a}} \subfigure[The reconfigurable reflect-array and the experiment testbed.]{
\includegraphics[width=2in]{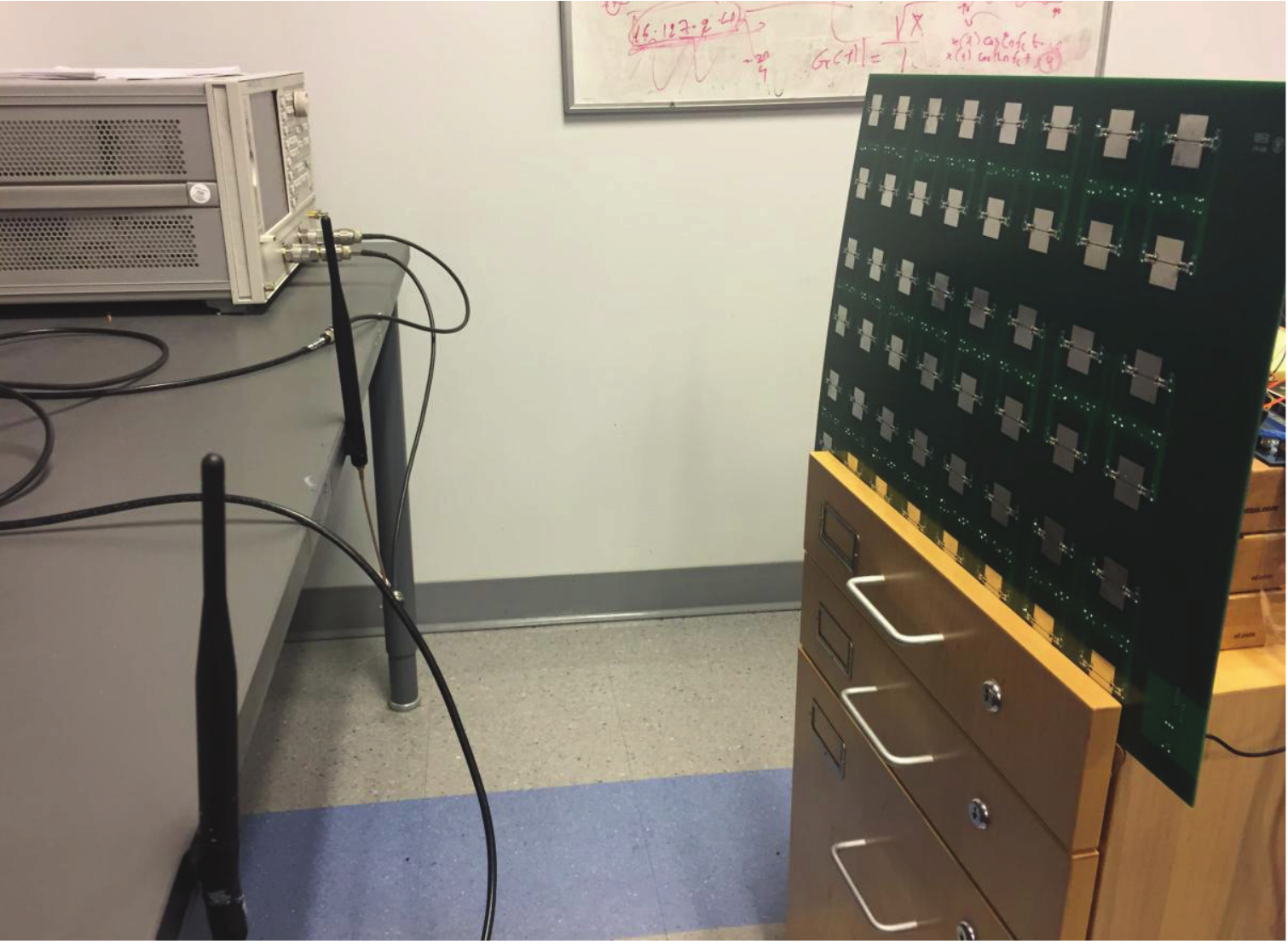}
\label{fig.01b}}
\caption{The proof-of-concept experiment for smart reflect-array.}\label{fig.01}
\end{figure}


\subsection{Experiment Results}

\begin{figure}
\centering
\subfigure[Tx1]{
\includegraphics[width=1.6in]{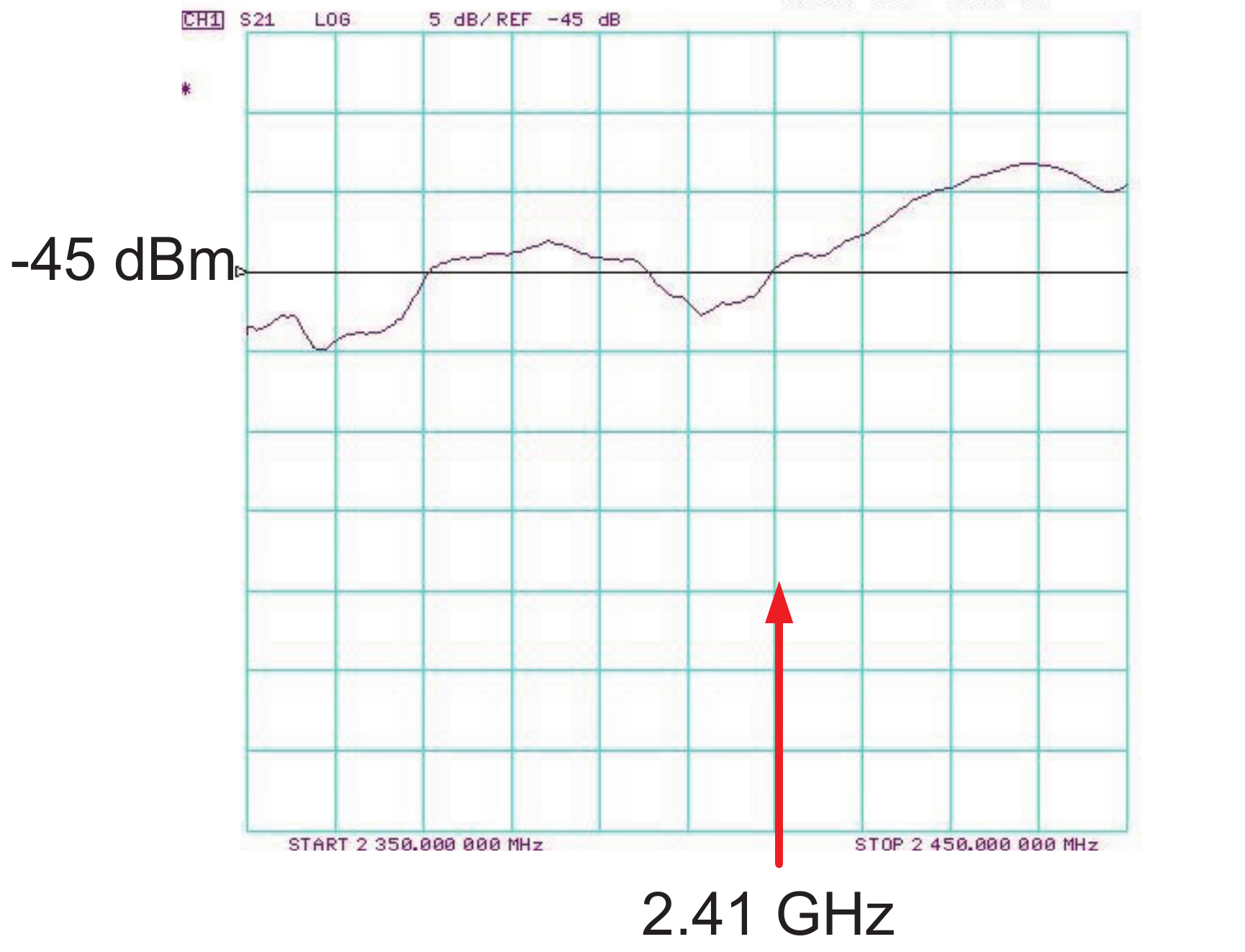}
\label{fig.11a}} \subfigure[Tx2]{
\includegraphics[width=1.6in]{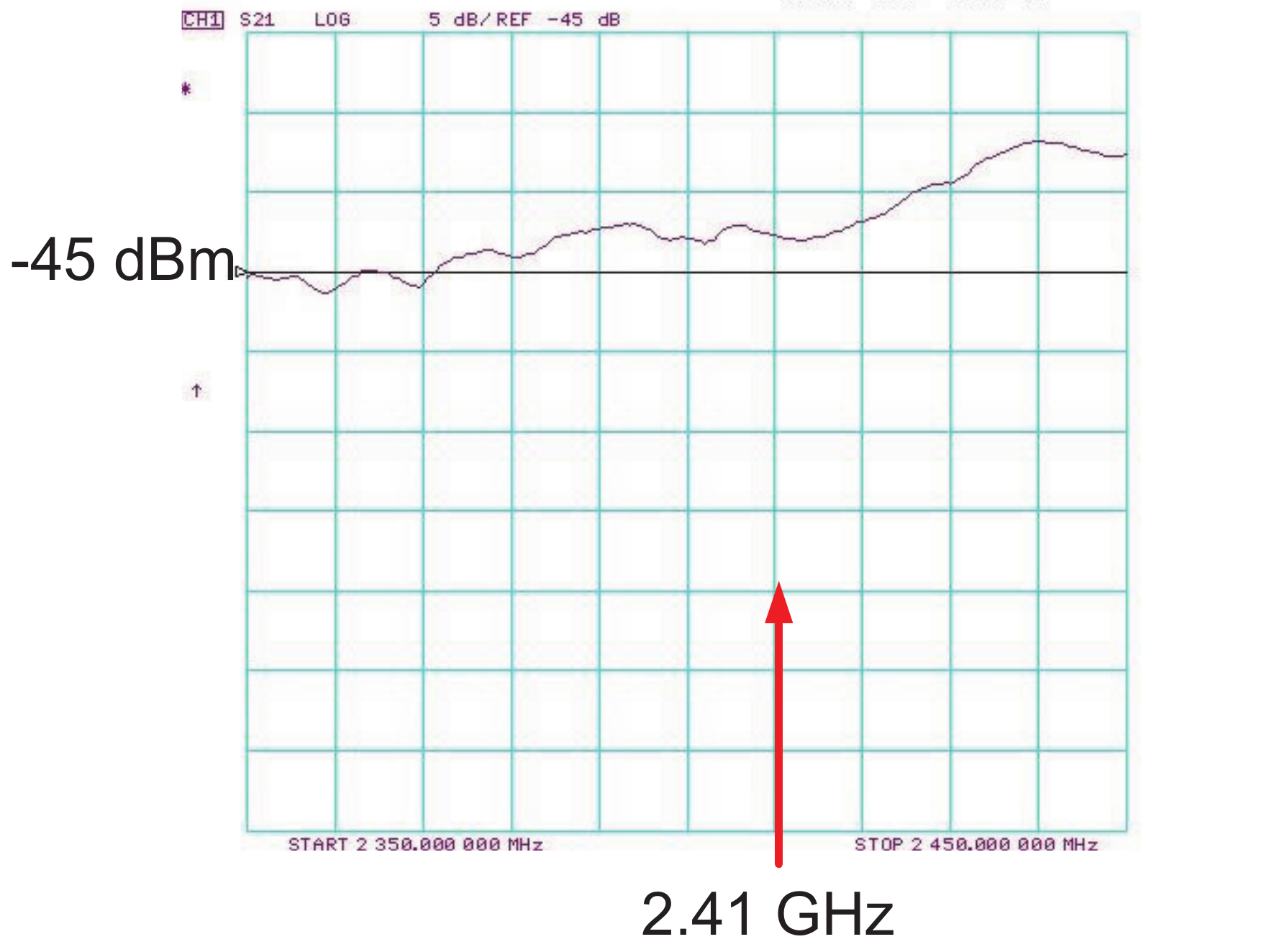}
\label{fig.11b}}
  \vspace{-5pt}
\caption{[Without reflect-array] The received signal strength from (a) Tx1 and (b) Tx2. }
  \vspace{-10pt}
\label{fig.11}
\end{figure}

\begin{figure}
\centering
\subfigure[Tx1]{
\includegraphics[width=1.6in]{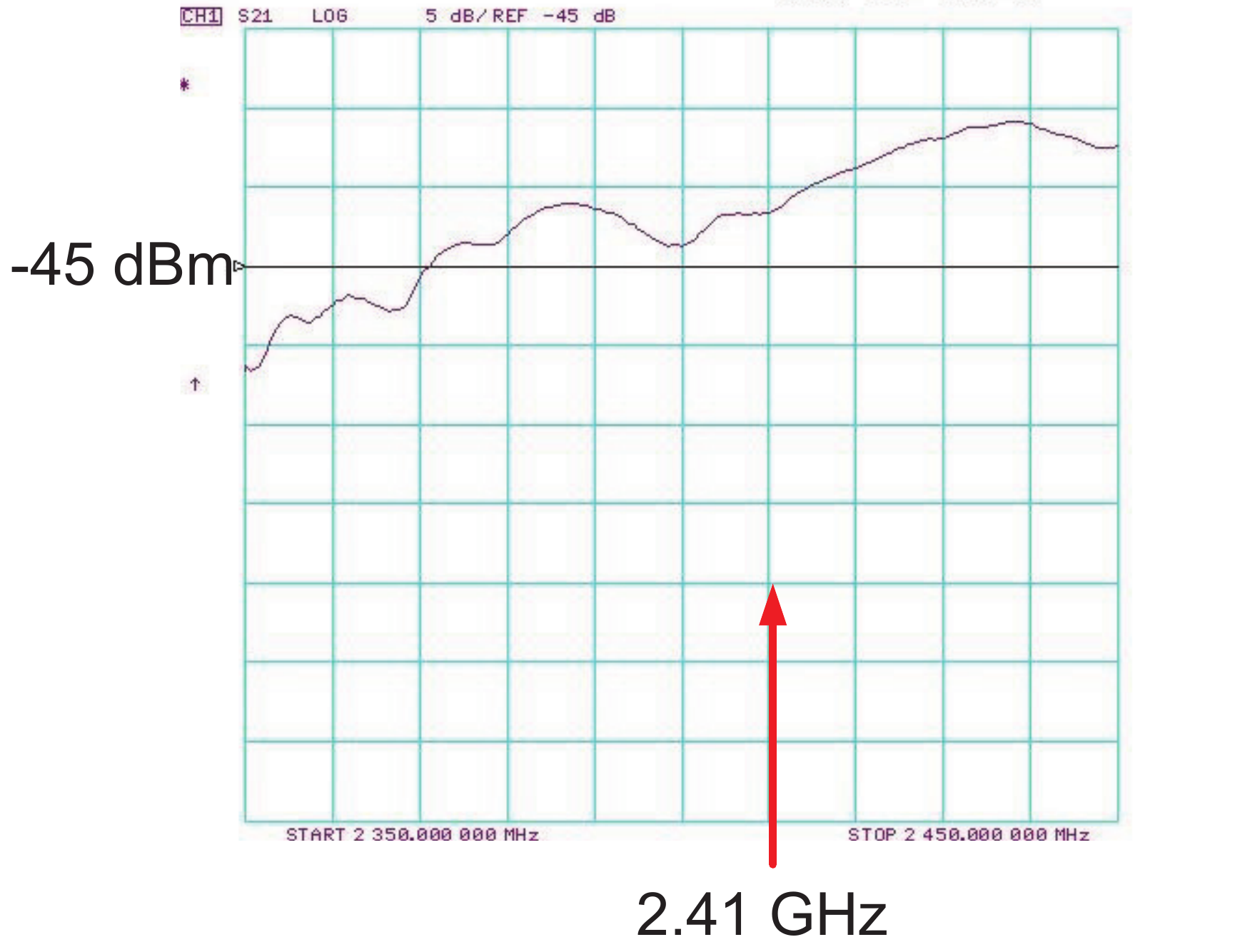}
\label{fig.12a}} \subfigure[Tx2]{
\includegraphics[width=1.6in]{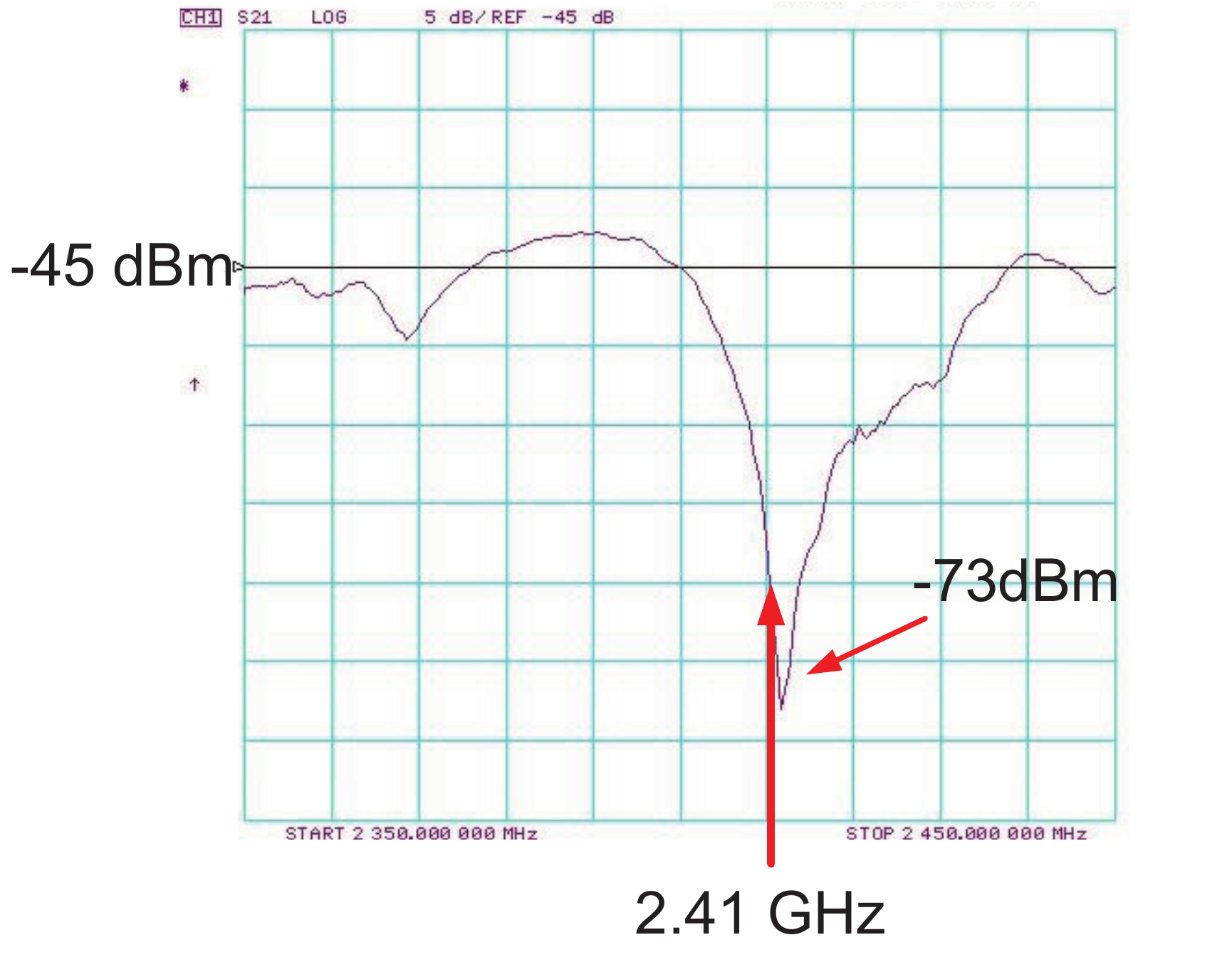}
\label{fig.12b}}
  \vspace{-5pt}
\caption{[With reflect-array] The received signal strength from (a) Tx1 and (b) Tx2.}
  \vspace{-10pt}
\label{fig.12}
\end{figure}


The experiment setup is illustrated in Fig. \ref{fig.01a}, where a receiver (Rx) is deployed $0.6$ $m$ in front of the reflect-array. A transmitting antenna (Tx1) working as the signal source is deployed $0.6$ $m$ away on left of the receiver. Another transmitting antenna (Tx2) is deployed as well to interfere the communication between Tx1 and Rx. The reflect-array consists of $6 \times 8 = 48$ reflectors and each reflector is controlled by a bias voltage to tune the varactors for changing the phase shift. Micro-controllers are used to give a control voltage to each reflector patch. An overview of experimental facilities is shown in Fig. \ref{fig.01b}.


During the experiment, Tx1 and Tx2 are transmitting signals along the whole spectrum and the frequency response at Rx is observed by spectrum analyzer. In Fig. \ref{fig.11}, we respectively measure the received signal strength from Tx1 and Tx2 without reflect-array nearby. Since the two transmission distances are the same ($0.6$ $m$), the received signal strengths are almost the same around -45 dBm. In Fig. \ref{fig.12}, by deploying the reflect-array and optimally tuning each reflector, the interference has been canceled to -73 dBm and the interference-plus-noise ratio (SINR) is increased to about 30 dB. In this way, the communication can be established between Tx1 and Rx by preventing them from the interference of Tx2.

\section{Increasing Transport Capacity by Smart Reflect-Array}
The experimental results prove the concept of the proposed spectrum sharing solution, where two simultaneous transmissions are considered. Furthermore, we expect the smart reflect-array can be used to simultaneously accommodate a large number of indoor wireless users with different services in the very limited spectrum bands. The RF spectrum utilization efficiency can be leveraged to a brand new height, which benefit all possible wireless systems and services. Thus, it is necessary to explore the capacity of spectrum sharing by considering multiple users in indoor environments. In this section, we first analyze the effect of reflect-array on the communication by developing accurate channel models. Based on the channel model, an upper bound on transport capacity is theoretically derived. Then, an achievable transport capacity is derived based on a practical reflect-array control algorithm. It should be noted that, the objective in this paper is to find the maximum possible transport network capacity. Therefore, we assume the network topology can be arbitrarily designed to achieve such capacity. Such network topology is commonly considered as arbitrary network \cite{definition}.

\vspace{10pt}
\subsection{Analysis of the Effect of Reflect-array}

\begin{figure}
  \centering
  \includegraphics[width=2.4in]{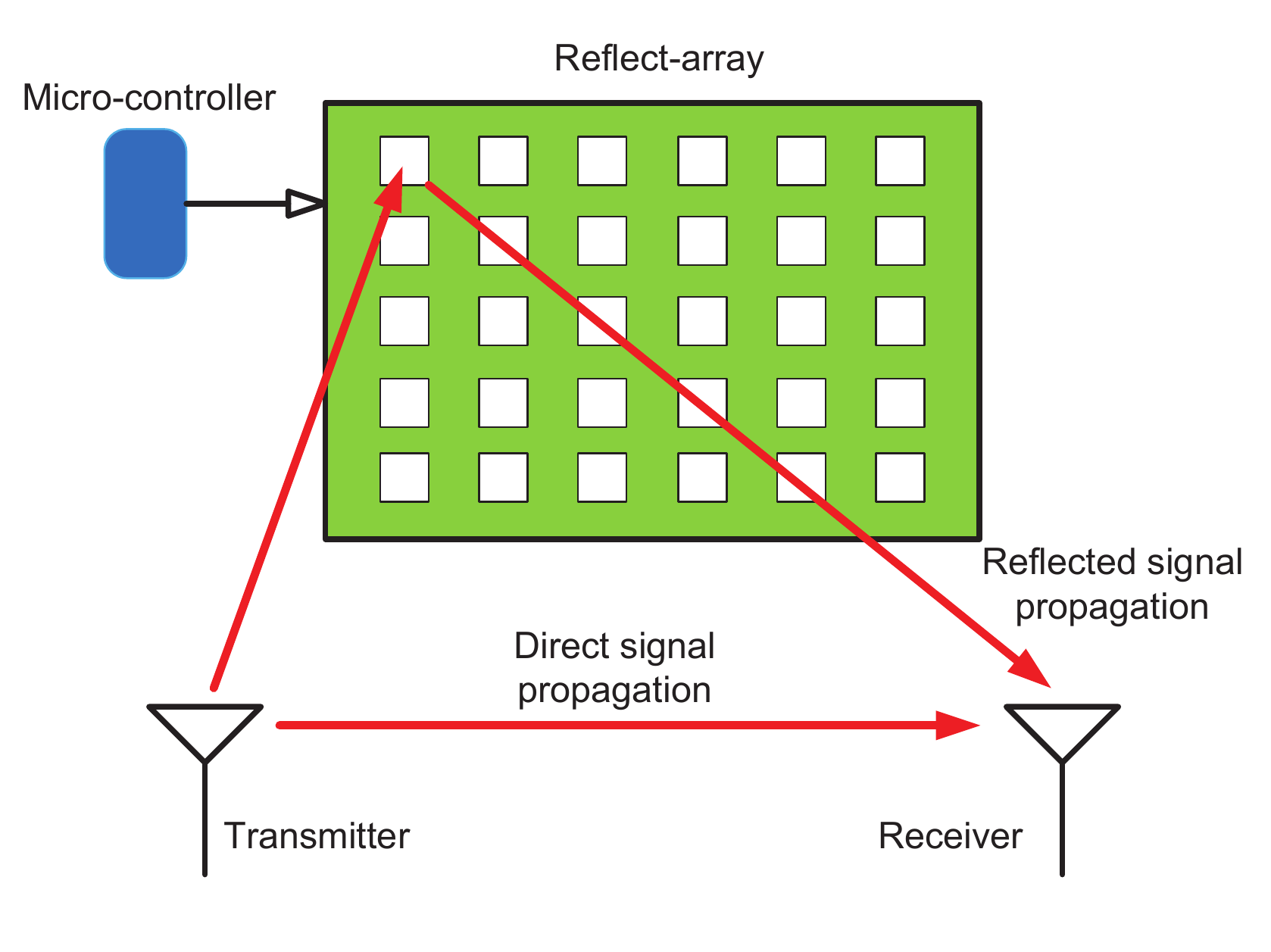}
    \vspace{-10pt}
  \caption{The channel influenced by the reflect-array.}
    \vspace{-5pt}
  \label{fig.1}
\end{figure}
\begin{figure}
  \centering
  \includegraphics[width=1.3in]{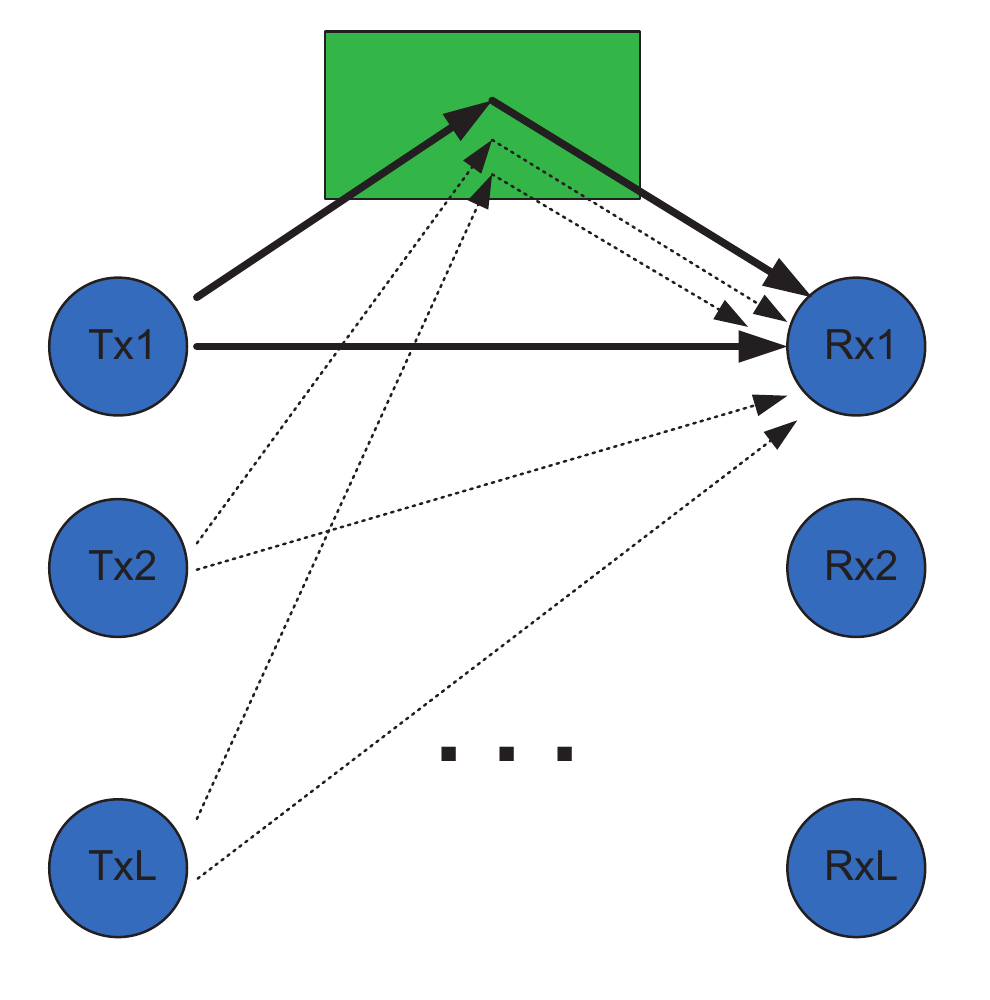}
    \vspace{-10pt}
  \caption{Multiple users served by the reflect-array}
    \vspace{-10pt}
  \label{fig.2}
\end{figure}

The influence from the reflect-array to the channel is shown as Fig. \ref{fig.1}. The signal received at the receiving side is the superposition of the direct signal and the signals reflected by patch antennas. Thus, the received signal strongly depends on the phases of the multi-path propagation.

Assuming that the signal transmitted on baseband is a train of raised cosines bearing BPSK symbols $m(t)$, the received signal from channel shown in Fig. \ref{fig.1} can be expressed in time domain:
\begin{equation}\label{received}
  r(t) = m(t) e^{j 2 \pi f_c} \left( a_0 e^{-j \theta_0} + \sum_{i=1}^{N} a_i e^{-j \left(\theta_i - \phi_i\right)} \right) + n \left(t\right)
\end{equation}
where $f_c$ is the operating frequency. $a_i$, $\theta_i$ are the attenuation and phase shift of the $i$-th path ($0$-th path corresponds to the LOS), respectively. $\phi_i$ is the phase induced by the $i$-th reflector for all $i=1, 2, ..., N$. $n(t)$ is the noise component of the received signal.

The reflect-array extends to serve in wireless networks accessed by multiple users. As shown in Fig. \ref{fig.2}, a number of $2L$ users are served by the reflect-array to establish $L$ pairs of communication. Not only the signal from the source will be reflected by the reflector but also the interferences from other transmitting nodes can get effect. Therefore, the received signal at a certain receiving node $l$ $(l=1, 2, ..., L)$ can be written as:
\begin{align}\label{multireceived}
\begin{split}
  & r_l(t)= m_l(t)e^{j 2\pi f_c t} \left( a_{l,l,0} e^{-\theta_{l,l,0}} + \sum^N_{i=1} a_{l,l,i} e^{- \left(\theta_{l,l,i} - \phi_i \right)} \right) \\
  & + \sum_{k \ne l} m_k(t) e^{j 2\pi f_c t} \left( a_{l,k,0} e^{-\theta_{l,k,0}} + \sum^N_{i=1} a_{l,k,i} e^{- \left(\theta_{l,k,i} - \phi_i \right)} \right) + n_l(t)
\end{split}
\end{align}
where the subscript $k$ indicates a transmitting node different from node $l$. For example, $a_{l,k,i}$ is the attenuation from transmitting node $l$ to receiving node $k$ through path $i$. The first term in (\ref{multireceived}) indicates the signal from the transmitting node while the second term is the signal from other $L-1$ interference sources. The phases of both signal and interference are controlled by induced phase $\phi_i$ on each reflector.

Since the received signal consists of the signal from the source, interferences and noise, whether the communication can be established or not depends on the signal to interference plus noise ratio (SINR). According to the received signal expressed in (\ref{multireceived}), the SINR for the $l$-th receiver can be derived as:
\begin{align}\label{SINR}
\begin{split}
  SINR_l= \frac{\rho_l^2(t) \left| \textbf{h}^H_{l,l} \textbf{v}_\phi \right|^2}{\sigma_l^2(t) + \sum_{k \ne l} \rho_k^2(t) \left|\textbf{h}^H_{l,k} \textbf{v}_\phi \right|^2}
\end{split}
\end{align}
where
\begin{align}\label{parameter}
\begin{split}
\textbf{h}_{l,k} \triangleq \left[ a_{l,k,0} e^{j \theta_{l,k,0}} ... \quad  a_{l,k,N} e^{j \theta_{l,k,N}}\right]^T, \quad \rho_l^2(t)\triangleq E\{\left|m_l(t)\right|^2\}, \\
\textbf{v}_\phi \triangleq [1 \quad e^{j \phi_1} ... \quad e^{j \phi_N}], \quad \quad \quad \quad \sigma_l^2(t) \triangleq E\{\left| n_l(t)\right|^2\}.
\end{split}
\end{align}

The effects from the reflect-array to the wireless networks are determined by the vector of the phase control $\textbf{v}_\phi$.

\subsection{An Upper Bound of Transport Capacity}
The focus of this subsection is to derive information theoretical upper bounds of transport capacity. According to \cite{definition}, the transport capacity of the network shown in Fig. \ref{fig.2} is defined as:
\begin{equation}\label{definition}
  C_T \triangleq sup \sum^{L}_{l=1} R \cdot d_{l,l,0}^\alpha
\end{equation}
where $d_{l,l,0}$ is the distance of the direct path between the $l$-th transmitter and $l$-th receiver. $\alpha$ is the rate of signal decay. $R$ is the feasible data rate if simultaneous reliable communication at rate $R$ is possible for all communication pairs. Since $R$ is usually set to be fixed in theoretical analysis, the upper bound of transport capacity is determined by $\sum^{L}_{l=1} d_{l,l}^\alpha$, which can be derived from the restriction of SINR:
\begin{equation}\label{restrictionSINR}
  SINR_l \geq \beta  \quad \quad (l=1, 2, ..., L)
\end{equation}
where $\beta$ is the threshold of SINR. According to (\ref{SINR}) and (\ref{restrictionSINR}), we have:
\begin{align}\label{SINRresult}
\begin{split}
 \frac{ \rho^2 \left| a_{l,l,0} + e^{j\theta_{l,l,0}} \sum^N_{i=1} a_{l,l,i}e^{\theta_{l,l,i}-\phi_i}\right|^2 }{\sigma^2+\sum_{k \in \Gamma} \rho^2 \left|a_{l,k,0} + e^{j \theta_{l,k,0}} \sum^N_{i=1} a_{l,k,i} e^{-\left(\theta_{l,k,i}-\phi_i\right)}\right|^2} \geq \frac{\beta + 1}{\beta}
\end{split}
\end{align}
where $\Gamma$ is the set of all receiving nodes. Thus, the denominator of the left side in (\ref{SINRresult}) indicates the noise plus the received power from the source and interferences. By considering the attenuation and phase shift with propagation distance, we have
\begin{equation}\label{set}
  a_{l,k,i}=\frac{1}{d_{l,k,i}^\alpha}, \quad \quad \theta_{l,k,i}= k_0 \Delta d_{l,k,i}
\end{equation}
where $k_0$ is the wave number. $d_{l,k,i}$ is the propagation length from transmitter $k$ to receiver $l$ reflected on the $i$-th element of reflect-array. $i=0$ means the direct path without reflection. $\Delta d_{l,k,i}$ is the difference of the propagation lengths that $\Delta d_{l,k,i}=d_{l,k,i}-d_{l,k,0}$. From (\ref{SINRresult}) and (\ref{set}), we can derive:
\begin{equation}\label{step1}
  d_{l,l,0}^\alpha < \frac{1}{ \left(\frac{\eta \sigma^2}{\rho^2} + \eta \left| I \right| \right)^{\frac{1}{2}}-\sum^N_{i=1} \frac{1}{d_{l,l,i}^\alpha} cos(k_0 \Delta d_{l,l,i} -\phi_i)}
\end{equation}
where $\eta=\frac{\beta + 1}{\beta}$ and $I$ is derived by:
\vspace{5pt}
\begin{align}\label{I}
\begin{split}
 I & =\sum_{k \in \Gamma} \left[ \frac{1}{d_{l,k,0}^\alpha} + \sum_{i=1}^N \frac{1}{d_{l,k,i}^\alpha} cos \left(k_0 \Delta d_{l,k,i}-\phi_i \right) \right]^2 \\
   & + \sum_{k \in \Gamma} \left[ \sum_{i=1}^N \frac{1}{d_{l,k,i}^\alpha} sin \left(k_0 \Delta d_{l,k,i}-\phi_i \right) \right]^2.
\end{split}
\end{align}
It is obvious that the upper bound is determined by the lower bound of the denominator of fraction in (\ref{step1}). Therefore, the parameter $I$ expressed in (\ref{I}) becomes the dominating factor of the transport capacity. By expanding the first term in (\ref{I}), we have:
\begin{align}\label{step2}
\begin{split}
&I=\sum_{k \in \Gamma} \left\{ d_{l,k,o}^{-2\alpha} + \sum_{i=1}^N 2 d_{l,k,0}^{-\alpha} d_{l,k,i}^{-\alpha} cos(k_0 \Delta d_{l,k,i} - \phi_i) + \right. \\
&\left. \left[\sum_{i=1}^N d_{d,k,i}^{-\alpha} cos(k_0 \Delta d_{l,k,i} - \phi_i)\right]^2  +\left[\sum_{i=1}^N d_{d,k,i}^{-\alpha} sin(k_0 \Delta d_{l,k,i} - \phi_i)\right]^2 \right\}  \\
 \quad &\geq \sum_{k \in \Gamma} \left\{ d_{l,k,o}^{-2\alpha} + \sum_{i=1}^N 2 d_{l,k,0}^{-\alpha} d_{l,k,i}^{-\alpha} cos(k_0 \Delta d_{l,k,i} - \phi_i) \right. \\
&\left. + \left[\sum_{i=1}^N d_{l,k,i}^{-\alpha} sin(k_0 \Delta d_{l,k,i} - \phi_i +\frac{\pi}{4})\right]^2 \right\}.
\end{split}
\end{align}
The inequation in (\ref{step2}) is achieved by the Arithmetic-Geometric average inequation that $x_1^2 + x_2^2 \geq \frac{1}{2}(x_1+x_2)^2$. Then, the result can be continuously derived as:
\begin{align}\label{step3}
\begin{split}
&\sum_{k \in \Gamma} \left\{ d_{l,k,o}^{-2\alpha} + \sum_{i=1}^N 2 d_{l,k,0}^{-\alpha} d_{l,k,i}^{-\alpha} cos(k_0 \Delta d_{l,k,i} - \phi_i) \right. \\
&\left. + \left[\sum_{i=1}^N d_{l,k,i}^{-\alpha} sin(k_0 \Delta d_{l,k,i} - \phi_i +\frac{\pi}{4})\right]^2 \right\}.\\
&\geq \sum_{k \in \Gamma} \left\{ \sum_{i=1}^N 2d_{l,k,0}^{-\alpha} d_{l,k,i}^{-\alpha} cos(k_0 \Delta d_{l,k,i} -\phi_i - \frac{\pi}{4})  \right.\\
&\left. +\sum_{i=1}^N 2d_{l,k,0}^{-\alpha} d_{l,k,i}^{-\alpha} cos(k_0 \Delta d_{l,k,i} -\phi_i) \right\}\\
&=2 \sqrt{\sqrt{2}+2} \sum_{k \in \Gamma} \sum_{i=1}^N d_{l,k,0}^{-\alpha} d_{l,k,i}^{-\alpha} cos(k_0 \Delta d_{l,k,i} -\phi_i -\frac{\pi}{8}).
\end{split}
\end{align}
In (\ref{step3}), we use the fact that $x_1^2 + x_2^2 \geq 2 x_1 x_2$ and trigonometric formula $cos A + cos B=2cos\frac{A+B}{2} \cdot cos\frac{A-B}{2}$.

In this derivation, we consider that the phase control $\phi_i \in [-\pi, \pi]$. The propagation distance $d_{l,k,i} \in [d_{min}, d_{max}]$, where $d_{min}$ and $d_{max}$ are respectively the minimum and maximum propagation length in the network. Obviously, $d_{min}$ and $d_{max}$ are determined by the geometry and dimension of the network. By substituting (\ref{step3}) into (\ref{step1}) and considering $L$ pairs of communications with a rate of $R$ for each, we have:
\begin{equation}\label{finalresult}
 C_T <\frac{RL}{ \left(\frac{\eta \sigma^2}{\rho^2}+ \frac{ \left(\sqrt{2}+2 \right) \eta N L}{d_{max}^{2 \alpha}}\right)^\frac{1}{2} -\frac{ N}{d_{min}^\alpha}}.
\end{equation}

\subsection{Achievable Bound of Transport Capacity for Arbitrary Networks}
The upper bound of transport capacity is theoretically calculated in above subsection. However, due to the restriction of the geometric of network, the propagation lengths $d_{l,k,i}$ cannot independently vary from $d_{min}$ to $d_{max}$. Thus, the upper bound becomes unachievable when a practical deployment of network is considered. In this subsection, we develop an algorithm to find out the achievable bound on transport capacity by considering the geometry of arbitrary networks.

\begin{figure}
\centering
\subfigure[Status n]{
\includegraphics[width=1.6in]{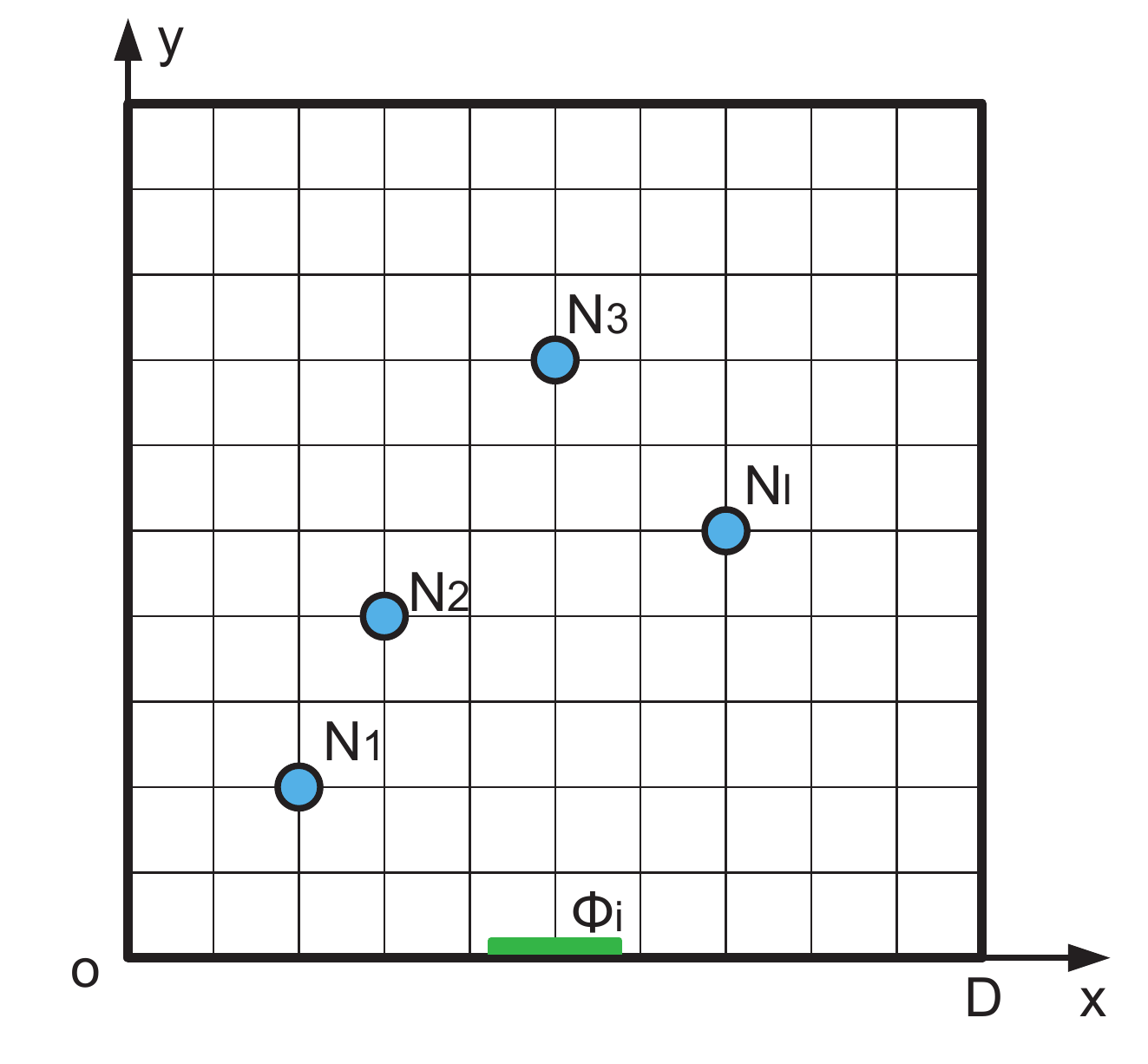}
\label{fig.3a}} \subfigure[Status n+1]{
\includegraphics[width=1.6in]{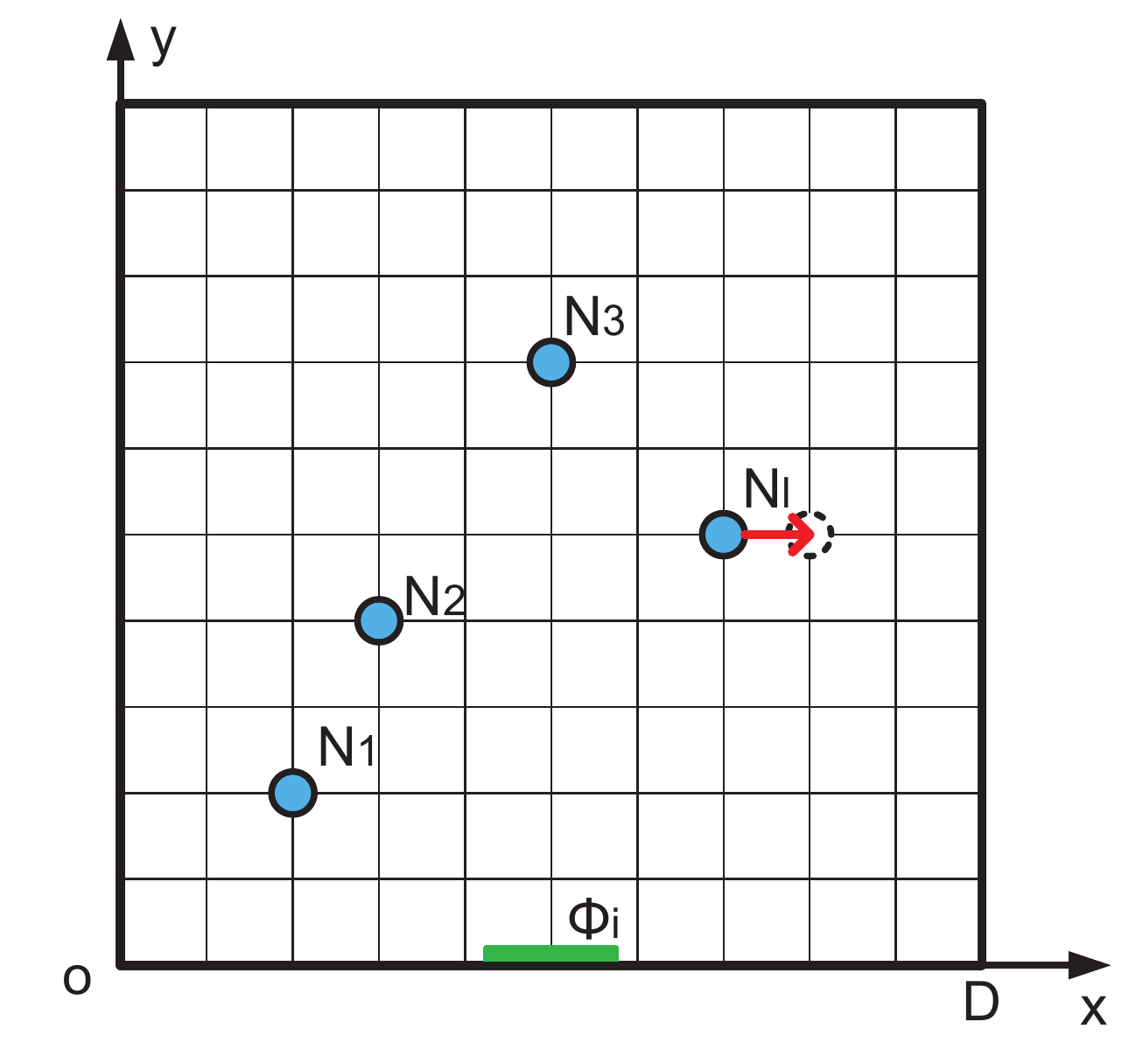}
\label{fig.3b}}
\caption{The statues of deployment of network.}\label{fig.3}
\end{figure}

As shown in Fig. \ref{fig.3}, a $D \times D$ $m^2$ square is divided by grid into small pixels with a number of $M \times M$. Therefore, the distance between two adjacent intersection points is determined by $\frac{D}{M}$ $m$. A reflect-array is located at $(\frac{D}{2},0)$ to serve the communication of the network. Then, we define a status vector to denote the positions of nodes shown in Fig. \ref{fig.3a}:
\begin{equation}\label{status1}
  \textbf{V}^{(n)}=\left\{(x^{(n)}_1, y^{(n)}_1), (x^{(n)}_2, y^{(n)}_2), ..., (x^{(n)}_{2L}, y^{(n)}_{2L})\right\}
\end{equation}
where the superscript $(n)$ indicates that the vector is for status $n$. $x^{(n)}_l$ and $y^{(n)}_l$ are the $x$ and $y$ positions of node $l$ in status $n$, respectively. After fixing the positions of the nodes, the phase control $\phi_i$ varies from $-\pi$ to $\pi$ to search for the maximum transport capacity using (\ref{step1}). For status $n$, the maximum transport capacity with optimal phase control $\phi_i^{(n)*}$ is denoted as $C_T^{(n)} (\mathbf{V}^{(n)}, \mathbf{\phi}_i^{(n)*})$. Then, shown as Fig. \ref{fig.3b}, we move one node in this network to get the next status $\textbf{V}^{(n+1)}$ by only changing
\begin{equation}\label{status2}
  x^{(n+1)}_l=x^{(n)}_l+\frac{D}{M}.
\end{equation}
Similarly, a maximum capacity for status $n+1$ can be achieved as $C_T^{(n+1)} (\mathbf{V}^{(n+1)}, \mathbf{\phi}_i^{(n+1)*})$. Therefore, by traversing all the combinations of deployment of nodes, the achievable bound on transport capacity can be found by:
\begin{equation}\label{maximumCT}
  C_{T\_max}=max \left\{C_T^{(1)}, C_T^{(2)}, ...,C_T^{(K)} \right\}
\end{equation}
where $K$ is the number of status in total. Detailed traversing for all the deployment status is shown in Algorithm 1.

\begin{algorithm}
\text{1.} \textbf{Input} $D$, $M$, $L$, $N$;\\
\text{2.} \textbf{for} $x_1 \gets 0$ \textbf{to} $D$ \textbf{step} $\frac{D}{M}$\\
\text{3.} \ \ \ \ \ \ $\cdot \cdot \cdot$\\
\text{4.} \ \ \ \ \ \ \ \ \ \  \textbf{for} $x_{2L} \gets 0$ \textbf{to} $D$ \textbf{step} $\frac{D}{M}$\\
\text{5.} \ \ \ \ \ \ \ \ \ \ \ \ \ \ \  \textbf{for} $y_{1} \gets 0$ \textbf{to} $D$ \textbf{step} $\frac{D}{M}$\\
\text{6.} \ \ \ \ \ \ \ \ \ \ \ \ \ \ \ \ \ \ \ \ \ \ \ $\cdot \cdot \cdot$\\
\text{7.} \ \ \ \ \ \ \ \ \ \ \ \ \ \ \ \ \ \ \ \ \ \ \ \ \ \ \  \textbf{for} $y_{2L} \gets 0$ \textbf{to} $D$ \textbf{step} $\frac{D}{M}$\\
\text{8.} \ \ \ \ \ \ \ \ \ \ \ \ \ \ \ \ \ \ \ \ \ \ \ \ \ \ \ \ \ \ \ \  \textbf{for} $\phi_{i} \gets -\pi$ \textbf{to} $\pi$ \textbf{step} $\frac{\pi}{180}$\\
\text{9.} \ \ \ \ \ \ \ \ \ \ \ \ \ \ \ \ \ \ \ \ \ \ \ \ \ \ \ \ \ \ \ \ \ \ \ \ \  \text{Calculate} $C_T$;\\
\text{11.} \ \ \ \ \ \ \ \ \ \ \ \ \ \ \ \ \ \ \ \ \ \ \ \ \ \ \ \ \ \ \ \ \ \ \ \  \textbf{if} $C_T > C_{T\_max}$\\
\text{12.} \ \ \ \ \ \ \ \ \ \ \ \ \ \ \ \ \ \ \ \ \ \ \ \ \ \ \ \ \ \ \ \ \ \ \ \ \ \ \  \textbf{then} $C_{T\_max} \gets C_T$;\\
\text{13.} \ \ \ \ \ \ \ \ \ \ \ \ \ \ \ \ \ \ \ \ \ \ \ \ \ \ \ \ \ \ \ \ \ \ \ \  \textbf{else} $C_T \gets 0$;\\
\text{14.} \ \ \ \ \ \ \ \ \ \ \ \ \ \ \ \ \ \ \ \ \ \ \ \ \ \ \ \ \ \ \ \ \ \ \ \  \textbf{end if}\\
\text{15.} \ \ \ \ \ \ \ \ \ \ \ \ \ \ \ \ \ \ \ \ \ \ \ \ \ \ \ \ \ \ \ \  \textbf{end for}\\
\text{16.} \ \ \ \ \ \ \ \ \ \ \ \ \ \ \ \ \ \ \ \ \ \ \ \ \ \ \  \textbf{end for}\\
\text{17.} \ \ \ \ \ \ \ \ \ \ \ \ \ \ \ \ \ \ \ \ \ \ \ $\cdot \cdot \cdot$\\
\text{18.} \ \ \ \ \ \ \ \ \ \ \ \ \ \ \  \textbf{end for}\\
\text{19.} \ \ \ \ \ \ \ \ \ \  \textbf{end for}\\
\text{20.} \ \ \ \ \ \ $\cdot \cdot \cdot$\\
\text{21.} \textbf{end for}\\
\text{22.} \textbf{Output} $C_{T\_max}$;\\
\caption{Finding the Maximum Transport Capacity}
\end{algorithm}

\section{Numerical analysis}

In this section, we first compare the upper bound of transport capacity mathematically derived in above section to the case that without reflect-arrays. Then, we evaluate the achievable bound by considering the geometry of networks.

\begin{figure}
\centering
\subfigure[Upper bound with number of communication pairs variation]{
\includegraphics[width=1.6in]{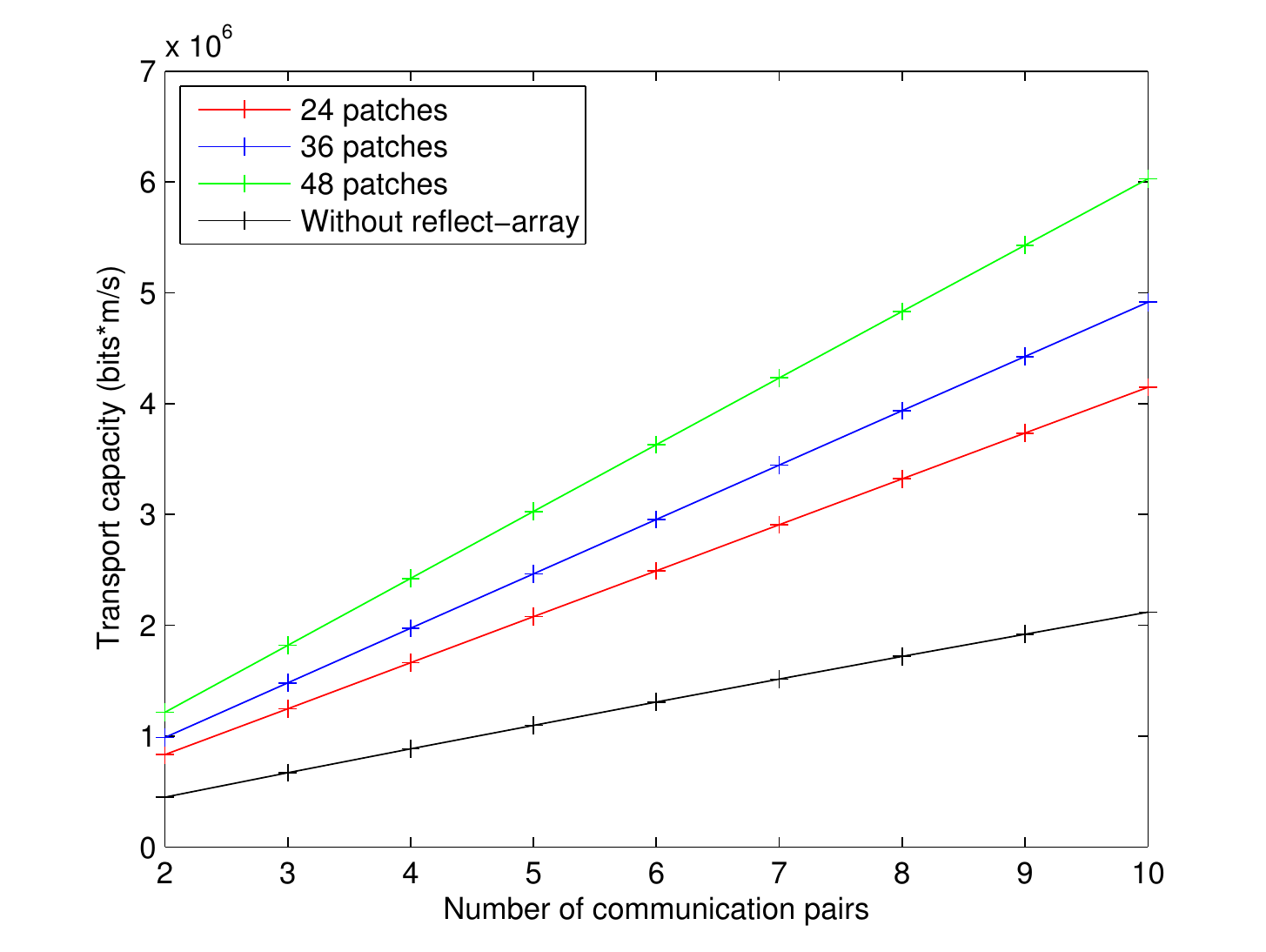}
\label{fig.4a}} \subfigure[Upper bound with edge length D variation]{
\includegraphics[width=1.6in]{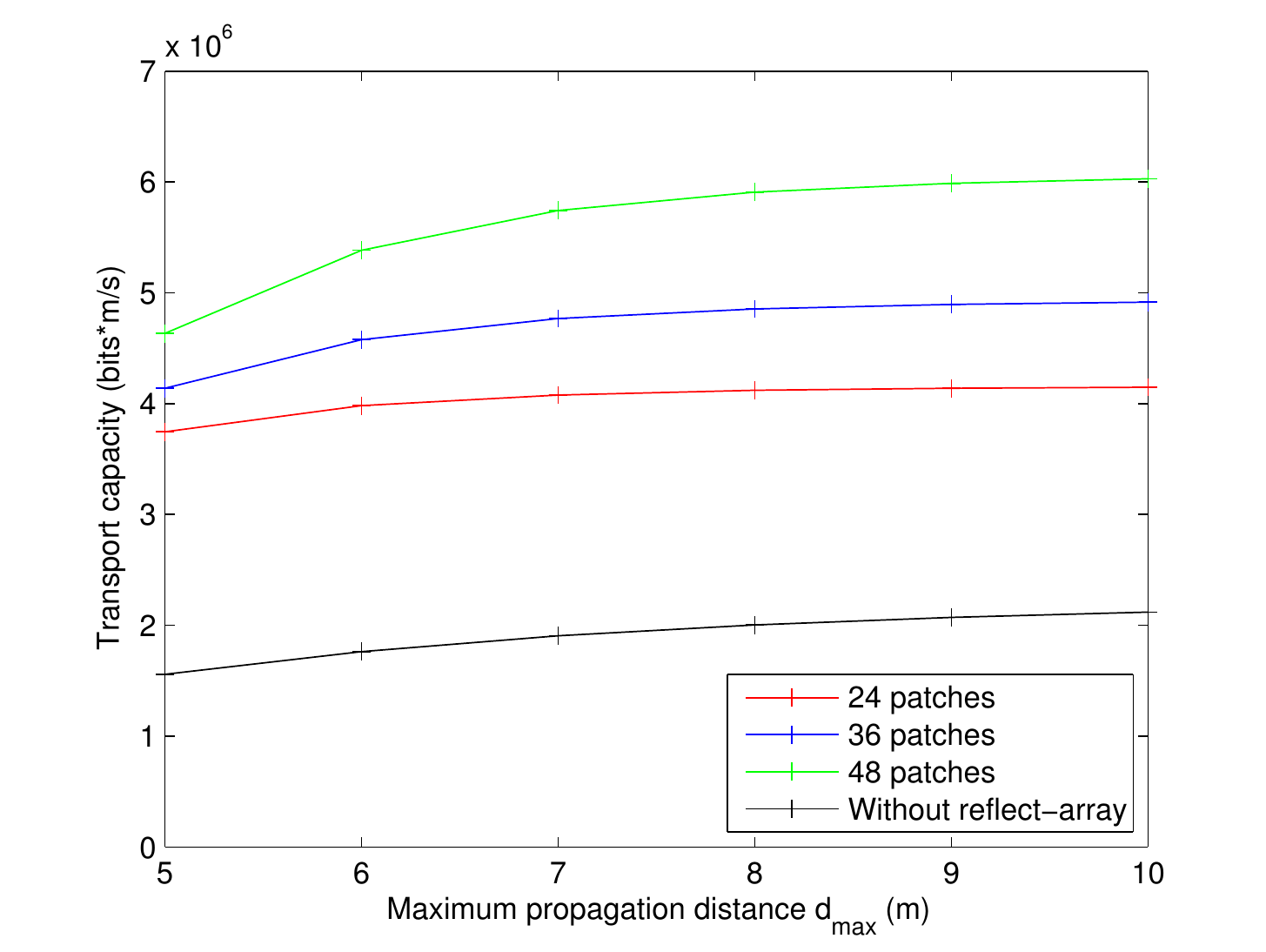}
\label{fig.4b}}
\caption{The upper bound of capacity.}\label{fig.4}
\end{figure}

Fig. \ref{fig.4a} shows an evaluation of the upper bound by varying the number of communication pairs. In this evaluation, communication pairs are considered to be deployed in a square room with an edge length of $D=10$ $m$. We use a transmitting power of $1$ mW for all transmitting nodes. The noise level is set to be $\sigma^2=-90$ dBm. The threshold of SINR $\beta=5$ dB. The rate of signal decay $\alpha=3$. The transmission rate $R=1\times 10^5$ $bits/s$. The red, blue and green curves respectively show the upper bound of transport capacity by using a reflect-array of 24, 36 and 48 patched reflectors. The black curve shows the upper bound of capacity without reflect-array. Obviously, the transport capacity can be improved by using a reflect-array. About $0.2 \times 10^6$ $bits \cdot m/s$ increase of capacity can be obtained by increasing the number of patches from 24 to 48. By increasing the number of communication pairs, the upper bounds become higher since the transport capacity of a network is defined as the summation of the capacities of all communications.

In Fig. \ref{fig.4b}, the edge length $D$ is varied from $5$ $m$ to $10$ $m$.  The transport capacity increases with edge length increasing since the interference nodes can be optimally deployed further away from the receiving nodes in a lager indoor space.

\begin{figure}
\centering
\subfigure[A network served by reflect-array]{
\includegraphics[width=1.4in]{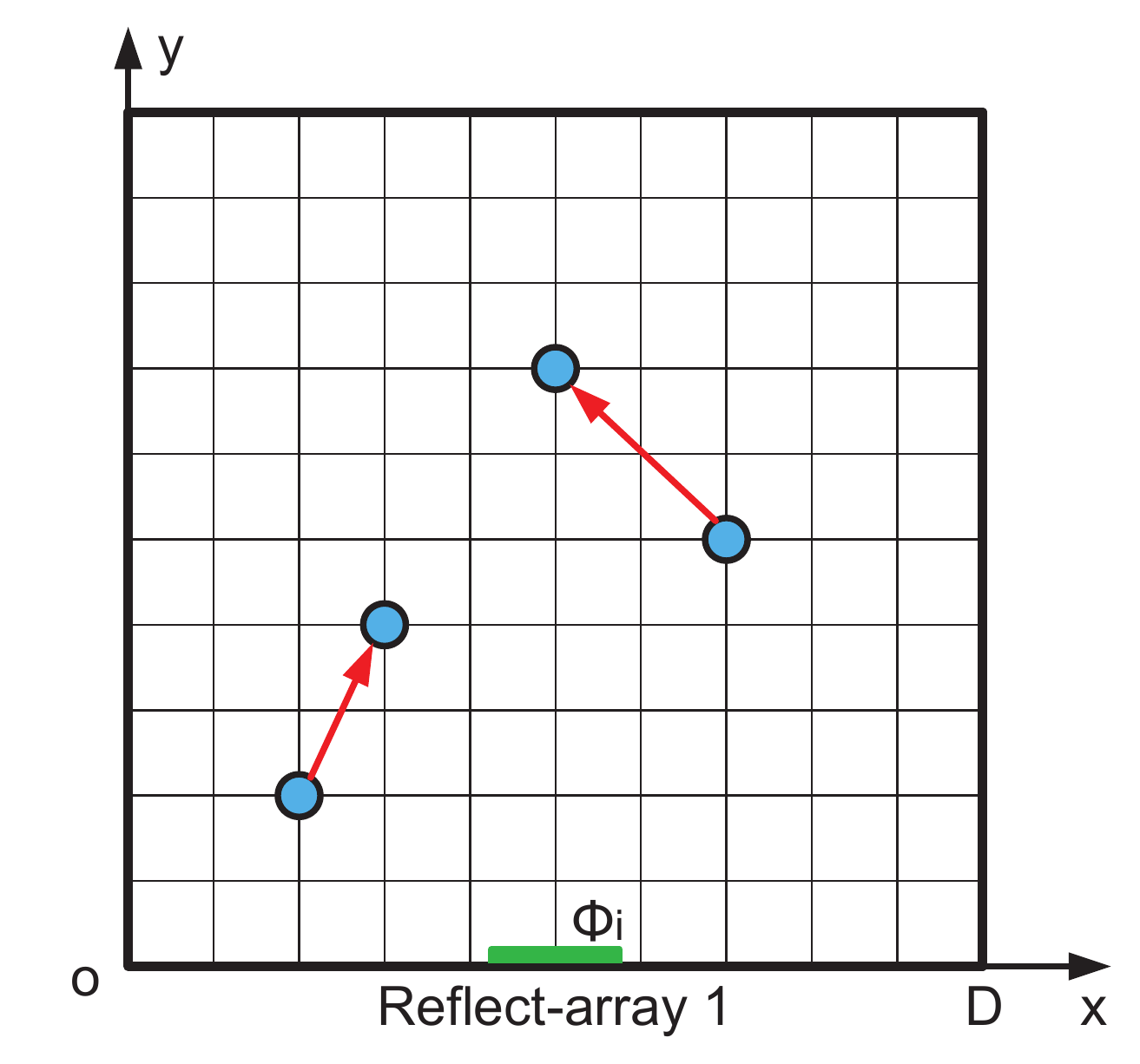}
\label{fig.5a}} \subfigure[Achievable bound on transport capacity]{
\includegraphics[width=1.8in]{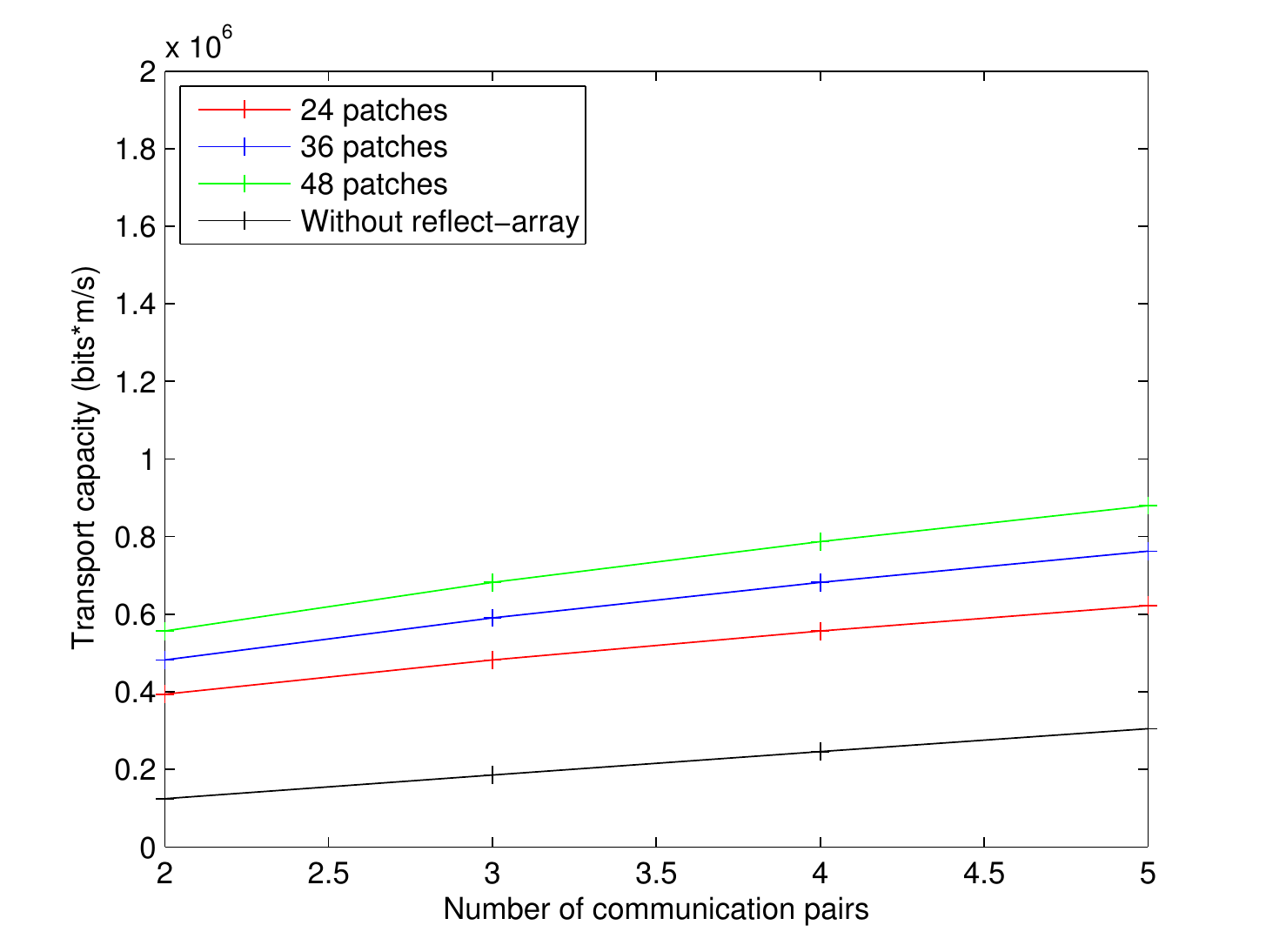}
\label{fig.5b}}
\caption{The achievable bound on transport capacity with number of communication pairs variation.}\label{fig.5}
\end{figure}

Achievable bounds on transport capacity are shown in Fig. \ref{fig.5}. As shown in Fig. \ref{fig.5a}, the reflect-array is deployed at $(\frac{D}{2},0)$. We divide the square into $10 \times 10$ pixels for the deployment of nodes. The phase control on reflect-array varies from $-\pi$ to $\pi$ with a step of $\frac{\pi}{180}$ to search for the optimal phases. Shown as the results in Fig. \ref{fig.5b}, the transport capacity can be improved by using a reflect-array with more patches.

\begin{figure}
\centering
\subfigure[A network served by two reflect-arrays]{
\includegraphics[width=1.4in]{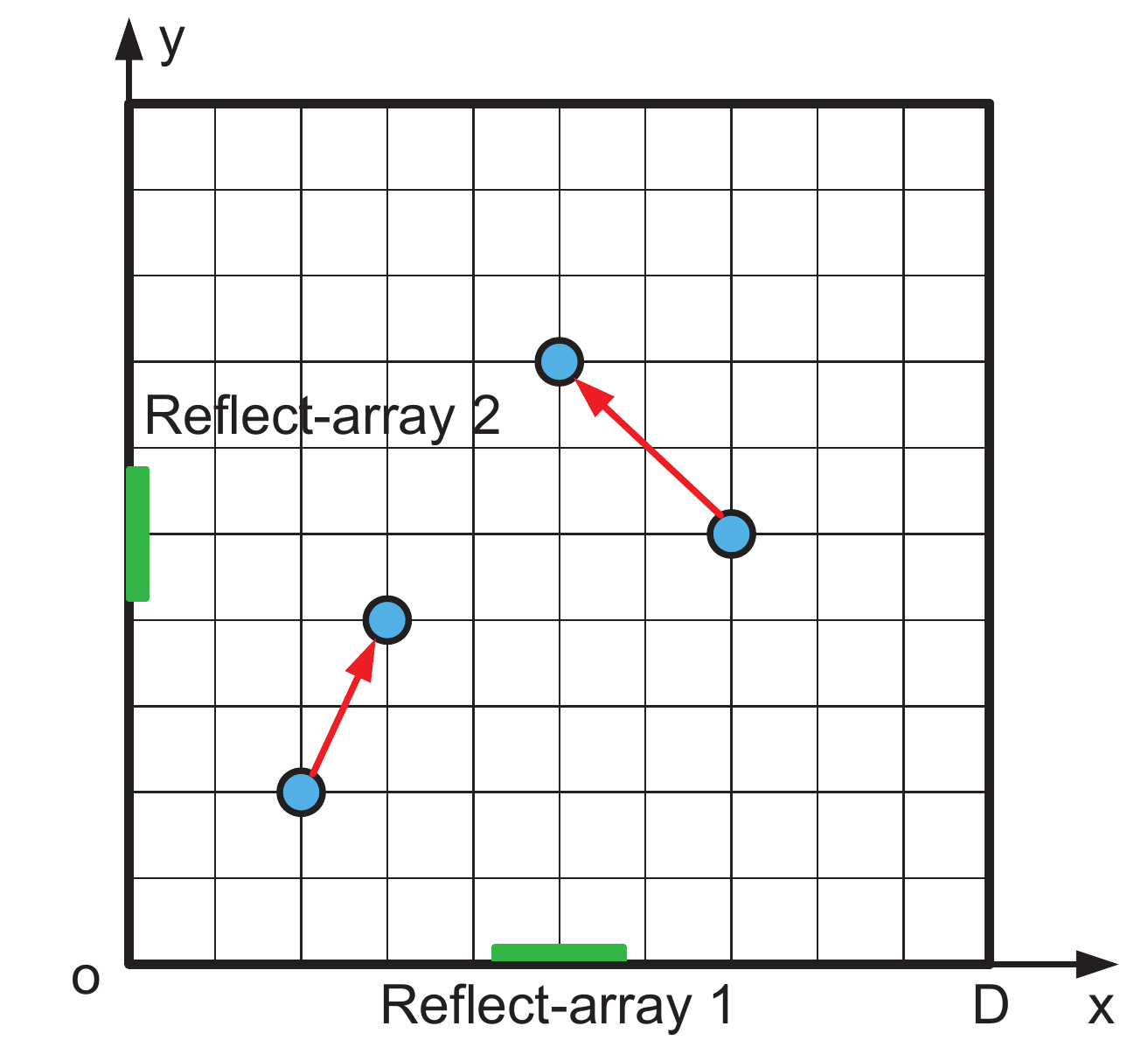}
\label{fig.7a}} \subfigure[Achievable bound on transport capacity]{
\includegraphics[width=1.8in]{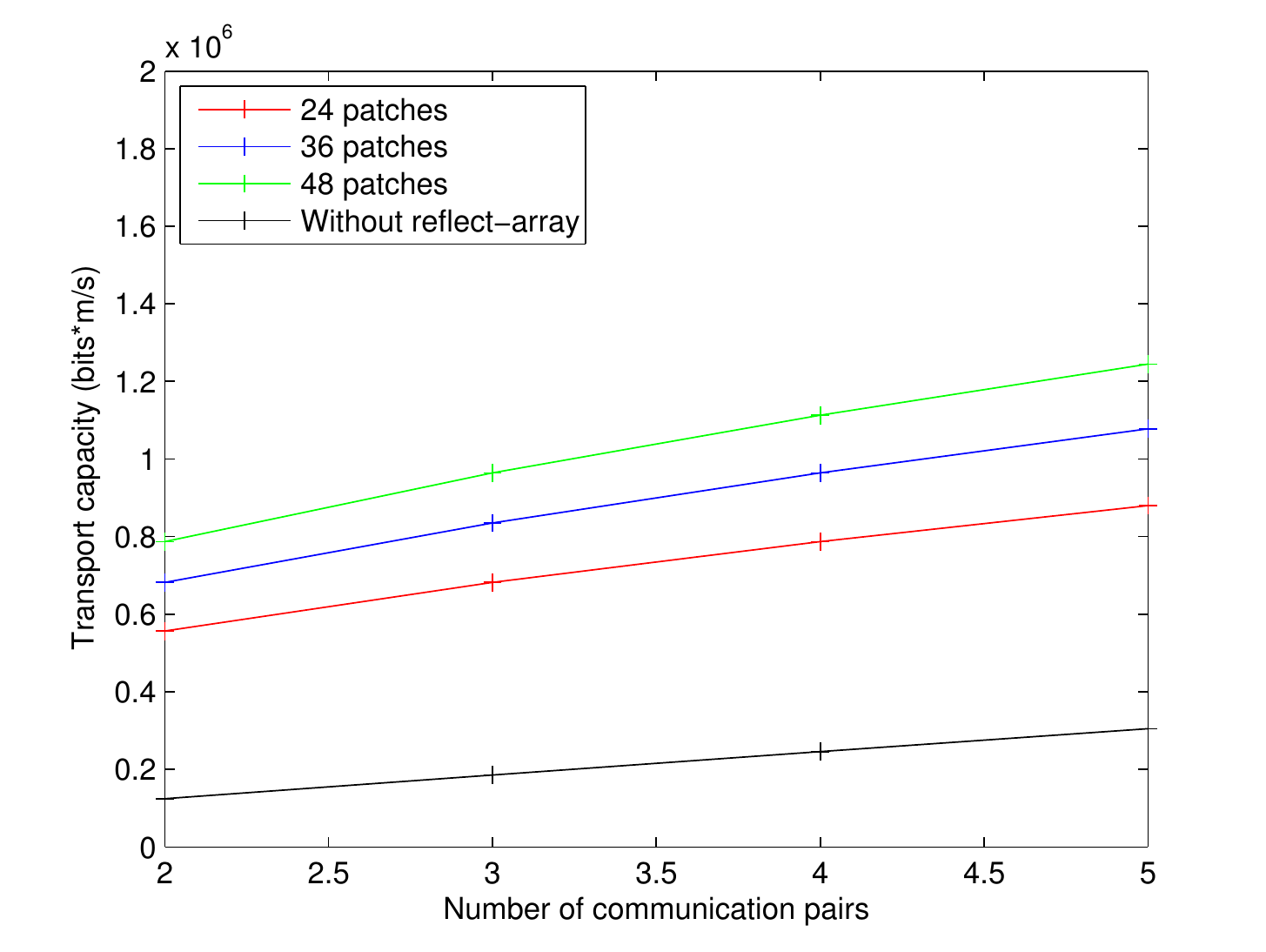}
\label{fig.7b}}
\caption{The achievable bound on transport capacity with number of communication pairs variation.(Two reflect-arrays)}\label{fig.7}
\end{figure}
\begin{figure}
\centering
\subfigure[A network served by three reflect-arrays]{
\includegraphics[width=1.4in]{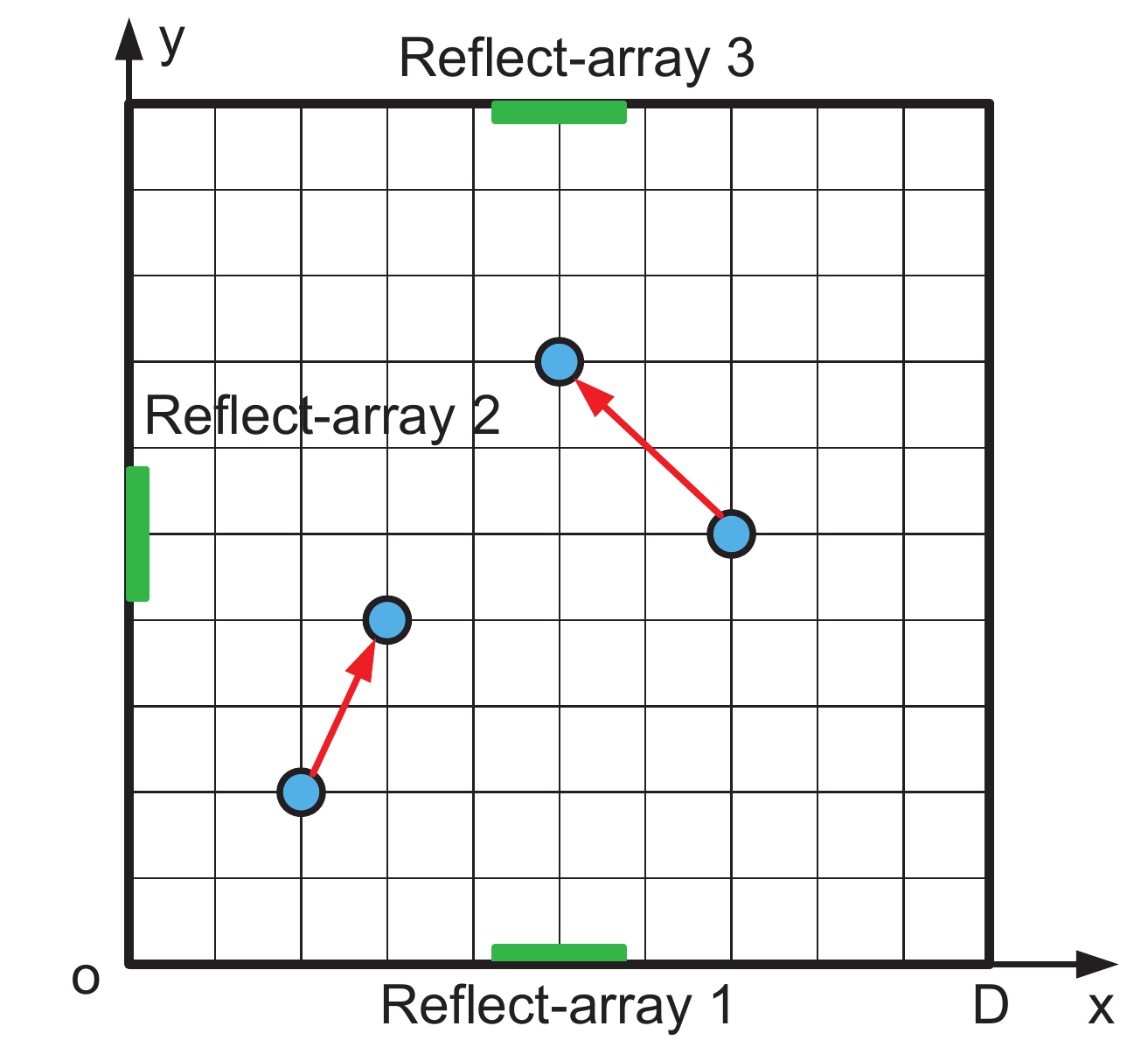}
\label{fig.8a}} \subfigure[Achievable bound on transport capacity]{
\includegraphics[width=1.8in]{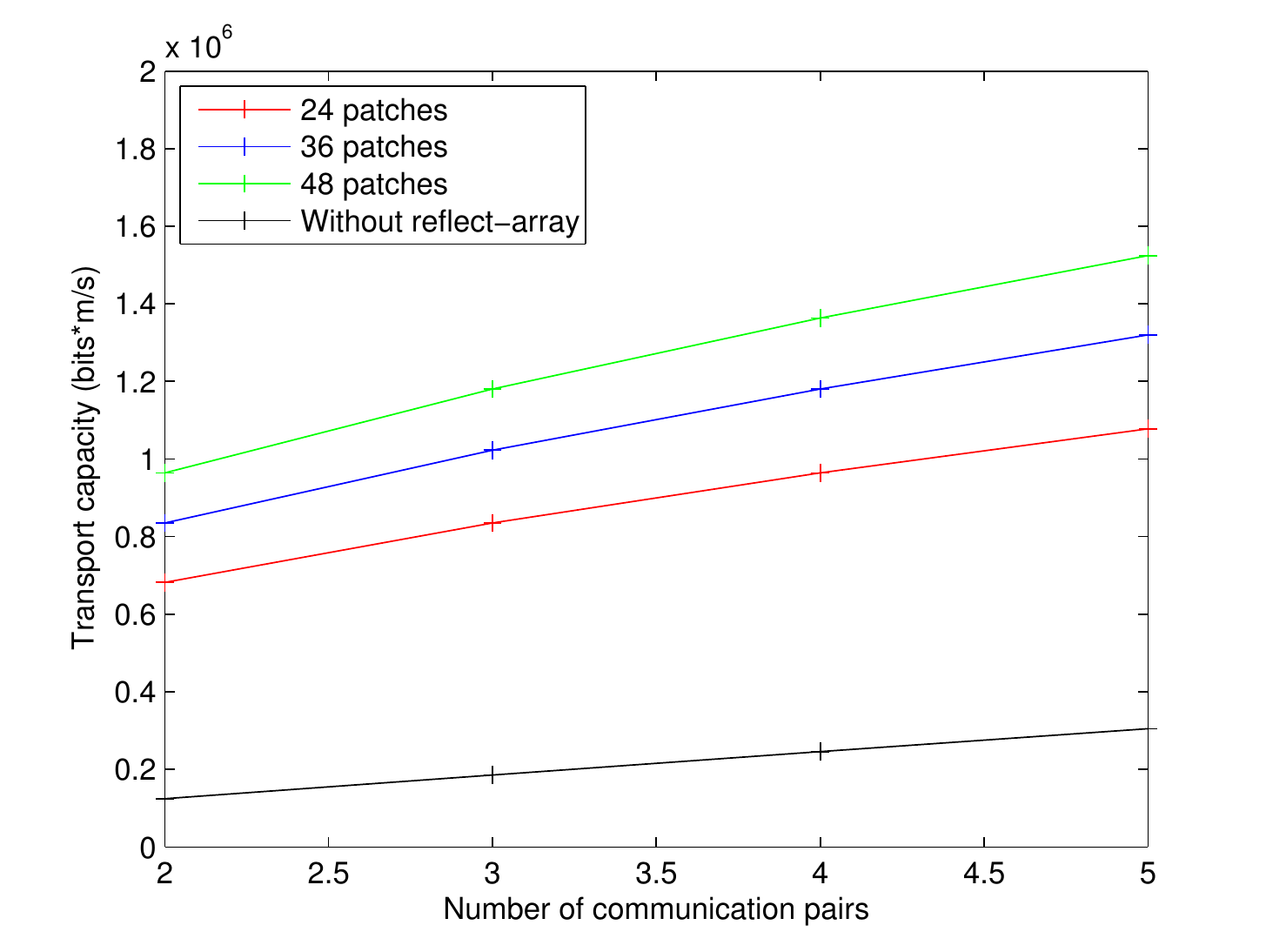}
\label{fig.8b}}
\caption{The achievable bound on transport capacity with number of communication pairs variation.(Three reflect-arrays)}\label{fig.8}
\end{figure}
\begin{figure}
\centering
\subfigure[A network served by four reflect-arrays]{
\includegraphics[width=1.4in]{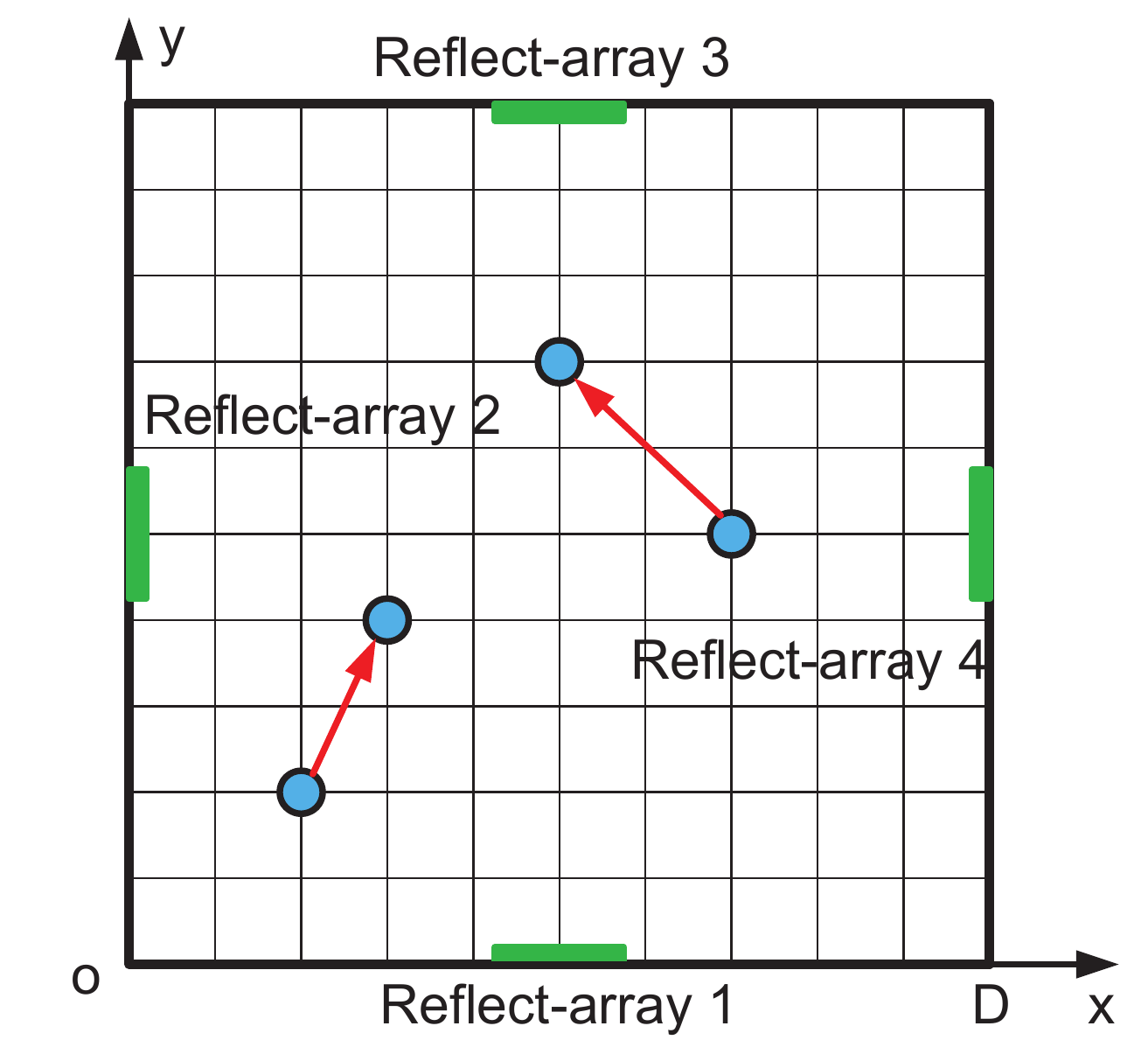}
\label{fig.9a}} \subfigure[Achievable bound on transport capacity]{
\includegraphics[width=1.8in]{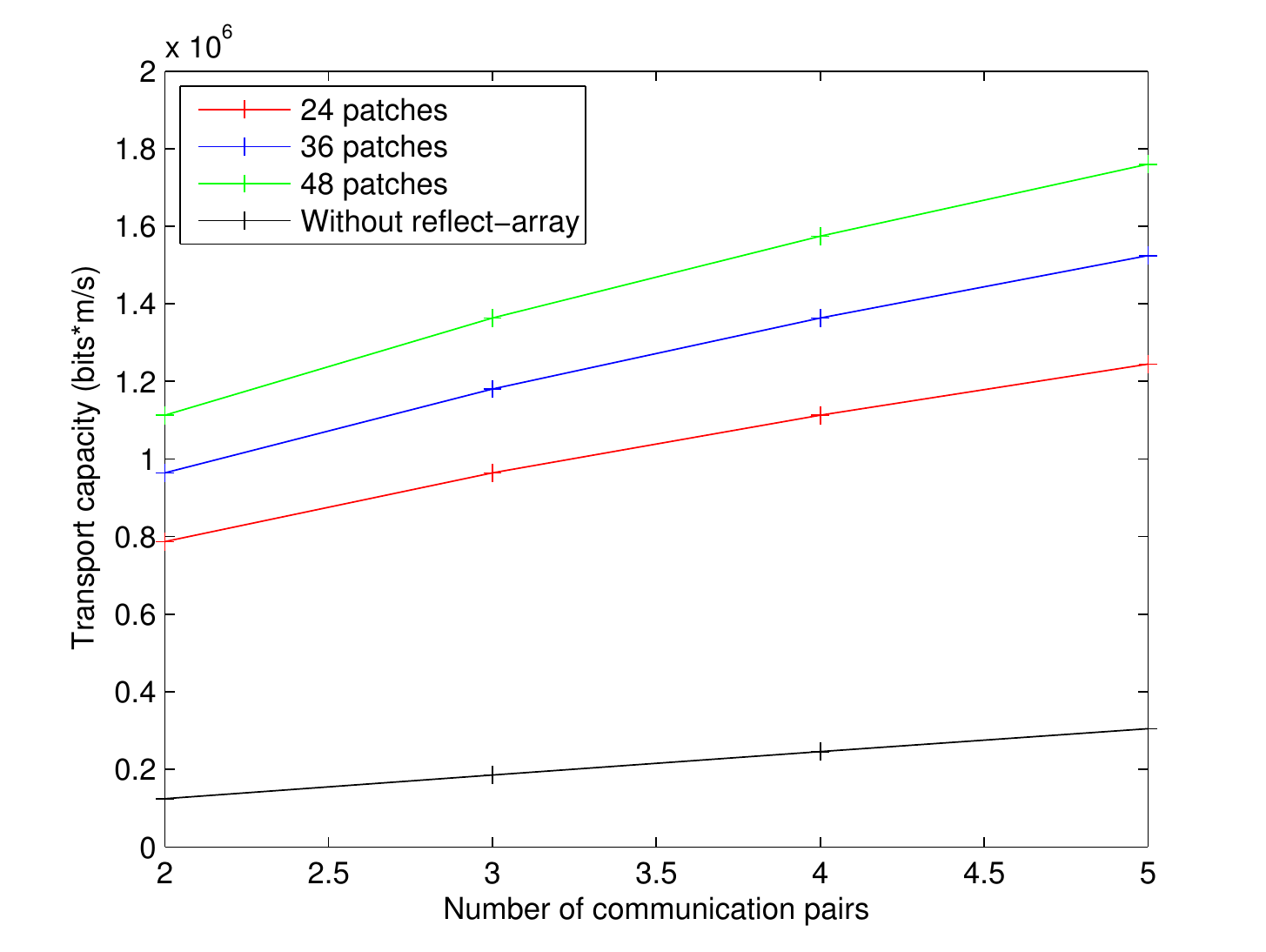}
\label{fig.9b}}
\caption{The achievable bound on transport capacity with number of communication pairs variation.(Four reflect-arrays)}\label{fig.9}
\end{figure}
\begin{figure}
\centering
\subfigure[The variation of edge length]{
\includegraphics[width=1.4in]{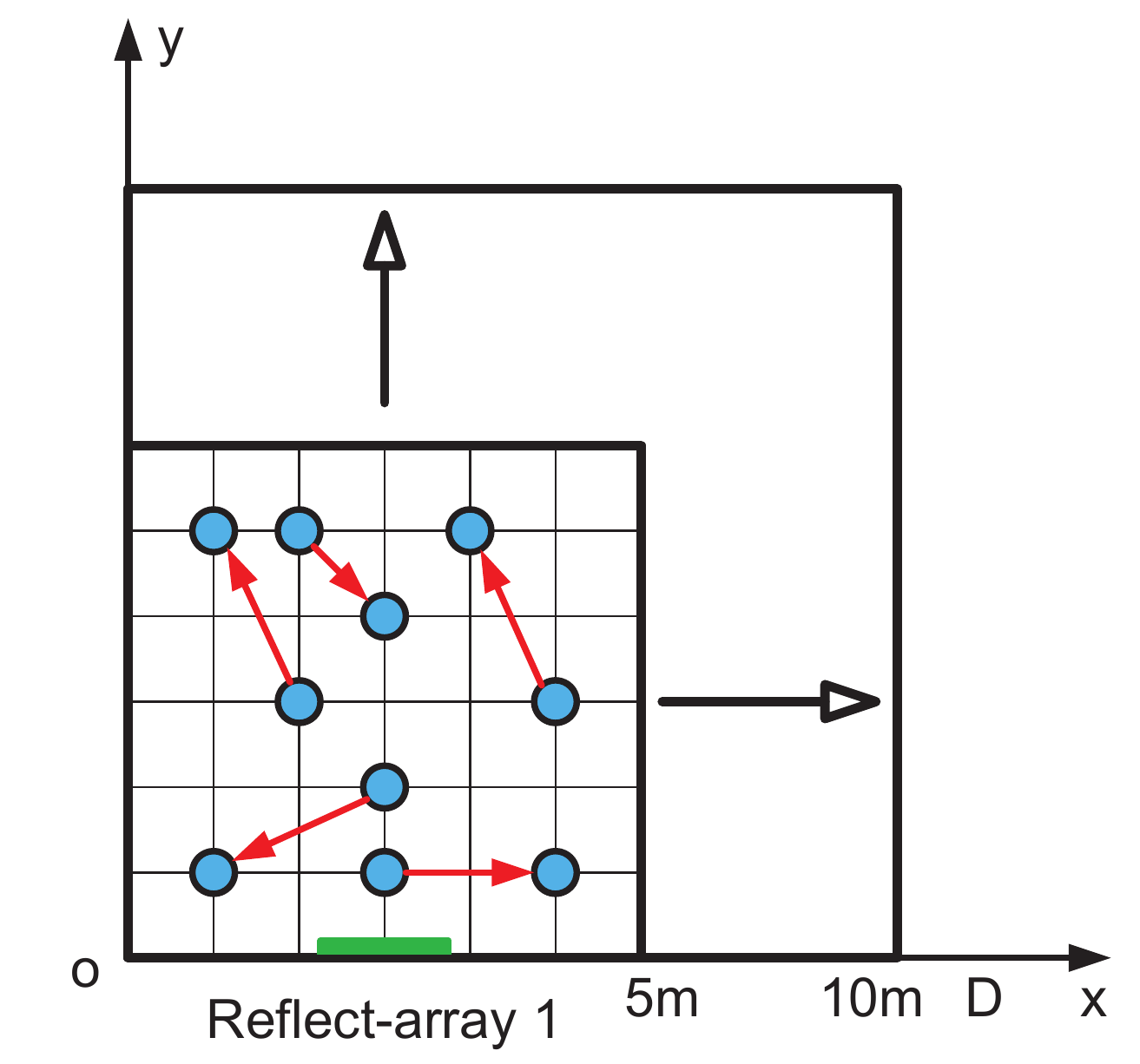}
\label{fig.10a}} \subfigure[Theoretical upper bound and achievable bound]{
\includegraphics[width=1.8in]{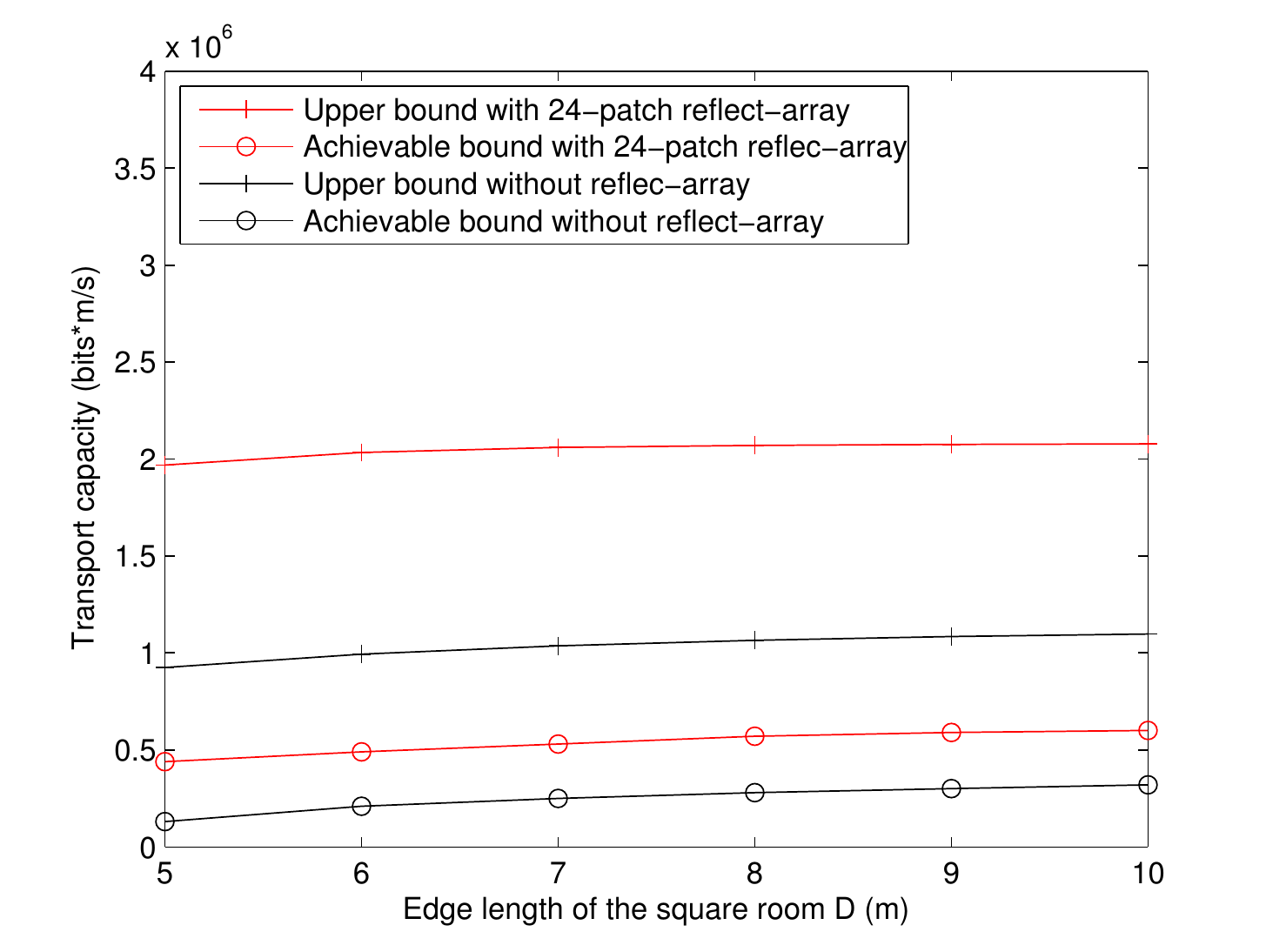}
\label{fig.10b}}
\caption{A comparison between upper bound and achievable bound on transport capacity.}\label{fig.10}
\end{figure}

In Fig. \ref{fig.7}, an extra reflect-array is deployed at $(0, \frac{D}{2})$ to serve the communication. Compared to the result shown in Fig. \ref{fig.5}, the capacity is further improved by $0.2 \times 10^6$ $bits \cdot m/s$ when two reflect-arrays are used. We observe the results by deploying the third and fourth reflect-array successively as shown in Fig. \ref{fig.8} and Fig. \ref{fig.9}. From \ref{fig.5b} to \ref{fig.9b}, the transport capacity obviously increases for about $0.5 \times 10^6$ $bits \cdot m/s$.

A comparison between the upper bound and the achievable bound is shown in Fig. \ref{fig.10}. As shown in Fig. \ref{fig.10a}, 5 pairs of nodes are deployed in a square room with one reflect-array located at $(\frac{D}{2}, 0)$. The edge length varies from $5$ $m$ to $10$ $m$. The results of using 24-patch reflect-array and without reflect-array are shown in Fig. \ref{fig.10b}. There exists a difference between two red/black curves since the theoretical upper bounds are derived by considering an ideal deployment of nodes which dose not exist in practical situations.

\section{Conclusion}
In this paper, we present a novel approach to improve the spectrum sharing capacity in indoor environments by using smart reflect-arrays. The feasibility of the approach has been verified by the experimental results. Theoretical derivation and algorithm have been developed to evaluate the effects of using reflect-array. Compared to the case without reflect-arrays, the numerical analysis shows a significant improvement on transport capacity when reflect-arrays are utilized.

Although the two-user experiments in this paper validates the feasibility of new spectrum sharing solution, our theoretical analysis shows that the smart reflect-array can be used for multiple simultaneous communications. Thus, new testbed for spectrum sharing with multi-users will be designed and implemented in the next step. Moreover, to accommodate multi-users and maintain robust communications in real time, optimal control algorithm will designed to configure the reflect-arrays according to the real-time spectrum usage in the indoor environment.

%

\end{document}